                        \newif\ifpaper \newif\ifPDF               %%
                        \newif\ifOUP \newif\ifboyscout            %%
                        \newif\ifdasbuch \newif\ifarticle         %%
                        \newif\ifsolutions

                        \newif\ifboyscout                         %%
                        \newif\ifpreparepdf                       %%
                        \preparepdftrue % hyperlinked pdf default %%

        \ifboyscout
\documentclass[final,1p,times,showpacs,hyperref]{elsarticle}
        \else
\documentclass[final,1p,times,showpacs,superscriptaddress]{elsarticle}
        \fi
\newcommand*{\capchi}{{\mathcal{X}}}
\usepackage{amssymb}
\usepackage{xcolor}
\journal{Physica D}

\begin{document}

\begin{frontmatter}



\title{Escape-rate response to noise of all amplitudes %and role %influence 
in leaky chaos}

%\author{Jeffrey M. Heninger$^{\mathrm{a}}$}
%\author{Domenico Lippolis$^{\mathrm{b,c},}$\footnote{domenico@ujs.edu.cn}}
%\author{Predrag Cvitanovi\'c$^{\mathrm{d}}$}
\author{Makoto Ohshika$^{\mathrm{a}}$, Domenico Lippolis$^{\mathrm{b},}$\footnote{domenico@ujs.edu.cn},  Akira Shudo$^{\mathrm{a}}$}
\address{$^\mathrm{a}$Department of Physics, Tokyo Metropolitan University, 
Minami-Osawa, Hachioji, Tokyo 192-0397, Japan}
\address{$^\mathrm{b}$Institute for Applied Systems Analysis,  Jiangsu University, Zhenjiang 212013, China}

\begin{abstract}
We study the effect of homogeneous noise
on the escape rate of %two-dimensional Anosov maps, 
%dynamical systems with small openings.
strongly  chaotic area-preserving maps with a small opening.
While in the noiseless dynamics 
%focusing especially on 
%the situation in which we put a sufficiently 
%small holes  %size 
%of similar shape to that of a typical region of a Markov partition. 
%The escape rate without noise is shown to be well described by a theoretical prediction 
the escape rate analytically depends on the instability of 
%containing the correction term associated with 
the shortest periodic orbit inside the hole,
adding noise overall enhances escape, which, however, exhibits a non-trivial response to the noise amplitude, featuring an initial plateau and a successive rapid growth up to 
a saturation value.     
%Increasing the noise amplitude does not    
%and the escape rate grows rapidly beyond the plateau. 
%The length of the plateau and the response to noise are found to depend 
 %on the instability of the shortest periodic orbit inside the hole. 
Numerical analysis is performed on cat maps with a hole, and the 
%We introduce a local model reproducing 
salient traits of the
response to noise of the escape rate are reproduced analytically by an approximate model.
%of the escape rate 
%subject to noise, and provide interpretations to numerical results. 

%Due to the high sensitivity to initial conditions, a chaotic system is expected to
%dramatically change the fractal geometry of its phase space, when the dynamical
%equations are even slightly modified, preventing any kind of perturbation approach.
%Consistently with the existing theory, we show an example of how the picture is improved by
%additive noise, that smoothens the stationary distribution,
%allowing perturbation theory to be applied in an otherwise unlikely scenario,
%to yield surprisingly accurate estimates of long-time averages.

\end{abstract}

\begin{keyword}
chaos \sep noise \sep escape \sep periodic orbits




\end{keyword}

\end{frontmatter}


\section{Introduction}

\label{sec:introduction}
Uniformly hyperbolic systems relax to a unique equilibrium state in a long time limit. In the case of area-preserving maps defined in a compact space, the equilibrium state is often given by the uniform Lebesgue measure. Thus, it is completely flat over the phase space. The rate of relaxation is exponential, reflecting the fact that the sub-leading eigenvalues of the evolution operator (Perron-Frobenius operator) of the system are discrete and isolated from others. Hence, they control the speed of relaxation. Na\"ively, such a feature was anticipated as a matter of course in the physics literature already after chaos was recognized. Rigorous mathematical results on the relaxation or the decay of correlations have been derived since the 1980s~\cite{Dorfle, Liverani95, Lustfeldetal96, Kaufmann96,dolgopyat1998decay,liverani2004contact}.  
The short-time dynamics reflects local structures in the phase space, so it is not necessarily uniform. On the other hand, the fact that the uniform Lebesgue measure gives the equilibrium state implies that the orbits explore the state space or phase space uniformly in a long time limit.

Recent studies~\cite{lippolis2021scarring,yoshida2021eigenfunctions} reveal that not everything proceeds homogeneously in space and time. Instead, the rate of correlation decay in closed systems depends on the position in the phase space, although the equilibrium state is uniform. This means that even relatively long-time dynamics reflect local phase space structures in the phase space. The existence of the strange eigenmode, that is the typically fractal pattern emerging in flow fields~\cite{aref2017frontiers} in the context of the so-called chaotic advection, could be regarded as a manifestation of inhomogeneous relaxation. It is observed that the residence time of the tracer particles on fluids depends on the position in the state space. Although the existence of  the strange eigenmode in the advection-diffusion equation has been rigorously proved in the system in an appropriate setting\cite{altmann2013leaking}, it is not yet clear enough what geometric structures of the underlying dynamical system the strange eigenmode reflects~\cite{aref2017frontiers}. 

The inhomogeneity in closed systems is carried over to open systems. In particular, leaky systems have been extensively studied for the past two decades, and substantial progress has been made, especially on the inhomogeneity of dynamics (Ref.~\cite{altmann2013leaking}). Leaky systems are obtained by punching a single or multiple holes in a closed system, and the orbits leak from the holes after a temporal exploration of the phase space. The leaky system requires a well-defined closed system that is used as a reference, and conversely, the closed system is usually obtained by taking the area of the leak to zero. In this sense, the 3-disk billiard system~\cite{chaosbook}, composed of three circular obstacles located without overlap, cannot be regarded as an example of leaky systems, although it is known to be a typical open system exhibiting chaotic scattering.

The escape rate is a natural figure of merit
to characterize the dynamics of leaky systems. In the presence of holes, the orbits launched in the phase space leak out from one of them. 
The escape rate is evaluated by counting the orbits that reach the holes per unit time interval. If the associated closed system is strongly chaotic, one would expect that the number of remaining orbits decreases exponentially fast because the process of hitting the holes must be Markovian. As a result, the escape rate is supposed to be proportional to the area of the leak. This statement can be more precisely specified in terms of the conditionally invariant density~\cite{demers2005escape,laitelbook}, 
and one should also note that the escape rate admits a general formula 
in terms of the trapped set of periodic orbits~\cite{dahlq99,altmann2009poincare}. 

As in the case of the position dependence of the relaxation rate in closed systems 
mentioned above, the escape rate depends on where one places the holes in the state space~
\cite{paar1997bursts,buljan2001many,schneider2002dynamics,altmann2004recurrence,bunimovich2007peeping,bunimovich2011place,afraimovich2010hole,keller2009rare,ferguson2012escape,bakhtin2011optimal,demers2005escape,altmann2009poincare,dettmann2011transmission,georgiou2012faster,demers2012behaviour,knight2012dependence,dettmann2012open,bunimovich2012fair,bunimovich2014improved,attarchi2020escape}. 
The position dependence would be one of the most important findings and thus has sparked renewed interest in the escape rate of expanding or hyperbolic systems, which had been assumed to be overall understood.

In particular, rigorous results for one-dimensional expanding linear maps show that significant corrections from short periodic orbits arise at the first nontrivial order in the expansion~%said ``as the next-order correction
\cite{keller2009rare,afraimovich2010hole,bunimovich2011place,bakhtin2011optimal,dettmann2012open,georgiou2012faster,ferguson2012escape,bunimovich2012fair,bunimovich2014improved}. 
The technique for their proofs is to use Markov partitions, and it has been extended to hyperbolic systems~\cite{attarchi2020escape,afraimovich2017escape}. 
The hole should also be chosen as an interval (region) of a Markov partition.

The exact formula is obtained in the limit of vanishing area of the leak, and the correction is expressed only in terms of the instability of the shortest periodic orbit. On the other hand, one should remember that the escape rate depends strongly on other parameters of the leaks, 
such as the orientation~\cite{schneider2002dynamics} and symmetry of the holes~\cite{dettmann2011transmission}. 
The change in the area of the leak can give rise to nontrivial behaviors~\cite{schneider2002dynamics,altmann2004recurrence,bunimovich2007peeping}. 
As found in \cite{bunimovich2011place,knight2012dependence,altmann2010noise,demers2012behaviour}, the parameter dependence of the escape rate can be even fractal. The issue is more subtle than it appears and should be explored carefully.

This report aims to describe the effect of noise on the escape rate in two-dimensional Anosov maps such as the cat map or its perturbed version. 
While studies of the response to noise of the diffusion coefficient and 
of the escape rate trace back to decades ago for low-dimensional maps~\cite{Frana91,Talk95,Reim96,CviDett1,CviDett2,CviDett3}, and dynamical systems of large spatial extension~\cite{Klages02,GaspardBook}, 
the present investigation is triggered by recent reports on the position dependence of the relaxation rate in closed systems from spatial profiles of eigenfunctions associated with sub-leading eigenvalues of the Perron-Frobenius operator, as well as of the effect of noise on the observed inhomogeneities~\cite{lippolis2021scarring,yoshida2021eigenfunctions}. As for the second-largest eigenvalues of the Perron-Frobenius operator, numerical calculations have shown that the noiseless limit behaves regularly, and so we would expect that the same is true for the spatial pattern of the corresponding eigenfunctions. Note that the role of noise or smoothing has been sought in proving the decay of correlations as well~\cite{blank2002ruelle}.  

Here, we address the same issue in the presence of a leak. As it is well known, the cat map is structurally stable~\cite{katok1997introduction}, that is robust against 
perturbations. Thus one may na\"ively expect that the escape rate for the system with sufficiently weak noise behaves similarly to the noiseless case. However, while in the deterministic system the long-time decay of the survival probability is governed by a fractal chaotic saddle~\cite{altmann2013leaking},  noise washes out the   
fractal structure of the non-attracting chaotic set, that, together with the subtleness of the position dependence mentioned above, makes
 the response of the escape rate to noise non-obvious. Therefore, our strategy here is to perform well-controlled numerical experiments in the noiseless limit and then tweak the noise amplitude to probe the response. 

To be specific, we prepare a sufficiently small single opening, whose shape mimics the Markov hole, which allows us to compare our outcomes to known results on the position dependence in the noiseless limit. Then we add uniformly distributed noise at each map iteration and 
evaluate the escape rate. There are several works on the effect of noise on the escape rate of leaky, strongly chaotic maps or billiards~\cite{altmann2010noise,altmann2012effect,bodai2013stochastic,da2018exploring}. 
We here examine the response profile of the escape as a function of the noise strength to see to what extent the features of deterministic dynamics resist noise, especially the role of periodic orbits with lower periods and smaller instability, and, on the other hand, how quickly the effect of noise becomes relevant and eventually dominant. Noise is in fact expected to govern the escape in its large-amplitude (compared to the area of the leak) limit. We also focus on the interplay between the noise strength and the area of the leak, by introducing a local model around the hole, and working out analytical 
estimates of the escape rate as a function of the noise amplitude.   

The organization of the paper is as follows. In section \ref{sec:Escape_rate}, we introduce the deterministic system to be studied here and derive a formula for the escape rate, to be used as a starting point for studying the effect of noise. The noiseless formula is consistent with the one derived in \cite{paar1997bursts,afraimovich2010hole,bunimovich2011place} for similarly strongly chaotic one-dimensional maps, 
and it is derived explicitly in subsection~\ref{subsec:derivation} for our system of choice. We then provide numerical evidence for the validity of the formula with the correction to the escape rate coming from periodic orbits. At the same time, it will be emphasized that taking the vanishing area of the leak and preparing the shape of the hole appropriately is crucial to improving the agreement. Section \ref{sec:escape_rate_with_noise} exposes the effect of noise on the escape rate. First, we present numerical results showing how the escape rate and its derivative vary as a function of the noise strength. We then introduce a local model around the opening, and derive analytical estimates for the escape rate, to be then compared with the numerical analysis. We also discuss the limit of small area of the leak. In section \ref{sec:summary}, we summarize our results and provide a brief outlook.

\section{Escape rate without noise} 
\label{sec:Escape_rate}

\subsection{Model and setting}
\label{sec:model_setting}

We here use the perturbed cat map 
% :\mathbb{T}^{2}\rightarrow\mathbb{T}^{2}$ 
defined on the torus $\mathbb{T}^{2} = [0,1) \times [0,1)$: 
\begin{equation}
\label
{eq:perturbed_cat_map}
F:
\left(\begin{array}{c}
\displaystyle 
X 
\vspace{1mm}
\\
\displaystyle 
Y
\end{array}
\right)
\mapsto
\left(\begin{array}{c}
\displaystyle 
X+Y-\frac{\epsilon}{\nu}\sin(2\pi\nu Y)
\vspace{1mm}
\\
\displaystyle 
X+2Y-\frac{\epsilon}{\nu}\sin(2\pi\nu Y) 
\end{array}\right)
~~
\mathrm{mod}\ 1. 
\end{equation}
The perturbation strength $\epsilon$ is assumed to be real and positive, and the perturbation frequency $\nu$ is a positive integer.
Note that in the limit $\epsilon =0$, the perturbed cat map $F$ is reduced to 
the cat map \cite{arnold1968ergodic}. 
It was shown that the perturbed cat map is topologically conjugate to the cat map 
if the condition $0 < \epsilon < (\sqrt{5} - 1) /4\sqrt{2}\pi \approx 0.069$
 is satisfied \cite{dematos1995quantization,dana2003chaotic}.
 
In order to evaluate the escape rate, it is important to locate the hole, denoted by $H$ hereafter. %for escape appropriately. 
As mentioned in the introduction, the shape, size, and orientation might affect the result. We here take the hole to be close to an element of the Markov partition. We refer to the formula derived in the paper \cite{bunimovich2012fair}, 
in which the fair-dice-like hyperbolic system has been rigorously examined to yield the formula for the escape rate by taking an element of the Markov partition as the hole. 
It is difficult, however, to construct the Markov partition analytically in the perturbed cat map $F$ except for the pure cat map case, {\it i.e.,} $\epsilon=0$. Thus, we here take the rhombus-shaped hole $H$, whose sides are respectively oriented in local stable and unstable directions at the center of the hole (see Fig.~\ref{fig:fig1}(a)). 

The other important setting for our subsequent analysis is that we place a periodic point at the center of the hole. 
More precisely, let $x_0=(X_{0}, Y_{0})$ be a periodic point of a period-$p$  %periodic %($\ge 1$)
orbit and $DF^{p}({x_0})$ be the associated Jacobian matrix, that is
\begin{equation}
DF^{p}({x_0}) = \prod_{k=0}^{p-1} J(x_k)\,, \hspace{1cm}					
J_{ij}(x) = \frac{\partial F_i}{\partial X_j}
\,.
\label{jac}
\end{equation}  
The directions of each side of the hole 
are given by the eigenvectors of the matrix $DF^{p}(x_0)$. %, and 
%put the point $\mathbf{x}_0$ at the center. 

In this setting, the formula derived in \cite{keller2009rare,afraimovich2010hole,bunimovich2011place,bakhtin2011optimal,ferguson2012escape,bunimovich2012fair} is expected to hold, so that we may identify the effect of noise as clearly as possible. If the area of the leak, denoted by $m(H)$,  is relatively large, the finite-size effect enters, which might be mixed with the effect of noise. 
In addition, in the presence of an explicit formula in the noiseless limit, we can develop an analytical argument under some simple assumptions, as the ones made below. 

%%%%%%%%%%%%%%%%%%%%%%%%%%%%%%%%%%%%%%
%%%%%%%%%%%%%%%%%%%%%%%%%%%%%%%%%%%%%%
\subsection{Formula for the escape rate}
\label{sec:formula}

When the area of the leak is sufficiently small, $m(H)\ll1$, the formula for the escape rate that we derive here 
is expressed as 
\begin{equation}
\label{eq:escape_rate}
\gamma(H)\simeq
m(H)\left(1-\frac{1}{\Lambda(p, x_0)}\right), 
%
%\begin{cases}
%m(H)\left(1-\cfrac{1}{\Lambda(p, (x_{c}, y_{c}))}\right)\qquad \mathrm{cat\ map}\\
%m(H)\left(1-\cfrac{1}{\Lambda_{\epsilon, \nu}(p, (x_{c}, y_{c}))}\right)\quad \mathrm{perturbed\ cat\ map}
%\end{cases}
\end{equation}
where $\Lambda(p, x_0)$ denotes 
the eigenvalue of the Jacobian matrix $DF^{p}({x_0})$ associated with 
the unstable direction, implying that $|\Lambda(p, x_0)| > 1$. 
We hereafter simply call $\Lambda(p, x_0)$ instability at 
the periodic point $x_0$, 
%\textcolor{red}{
and use the notation $\Lambda_p$ 
for the sake of brevity. 
%}
Obviously, the instability along the itinerary of a periodic orbit $F^{p}(x_0)$ 
is the same at any periodic point of the same orbit. 
Equation~(\ref{eq:escape_rate}) was first surmised in~ \cite{paar2000bursts}, and
therein tested on the H\'enon and Baker's map. In the next section we present a 
full derivation for the (unperturbed and perturbed) cat maps.
%The derivation of the formula (\ref{eq:escape_rate}) 
%is given in Appendix \ref{app:derivation_escape_rate}. 

%%%%%%%%%%%%%%%%%%%%%%%%%%%%%%%%%%%%%%%%
%%%%%%%%%%%%%%%       Fig. 1        %%%%%%%%%%%%%%%%
%%%%%%%%%%%%%%%%%%%%%%%%%%%%%%%%%%%%%%%%
\begin{figure}[ht]
\begin{center}
\begin{minipage}{1\hsize}
\hspace{5mm}
\includegraphics[width = 7.0cm,bb= 0 0 461 346]{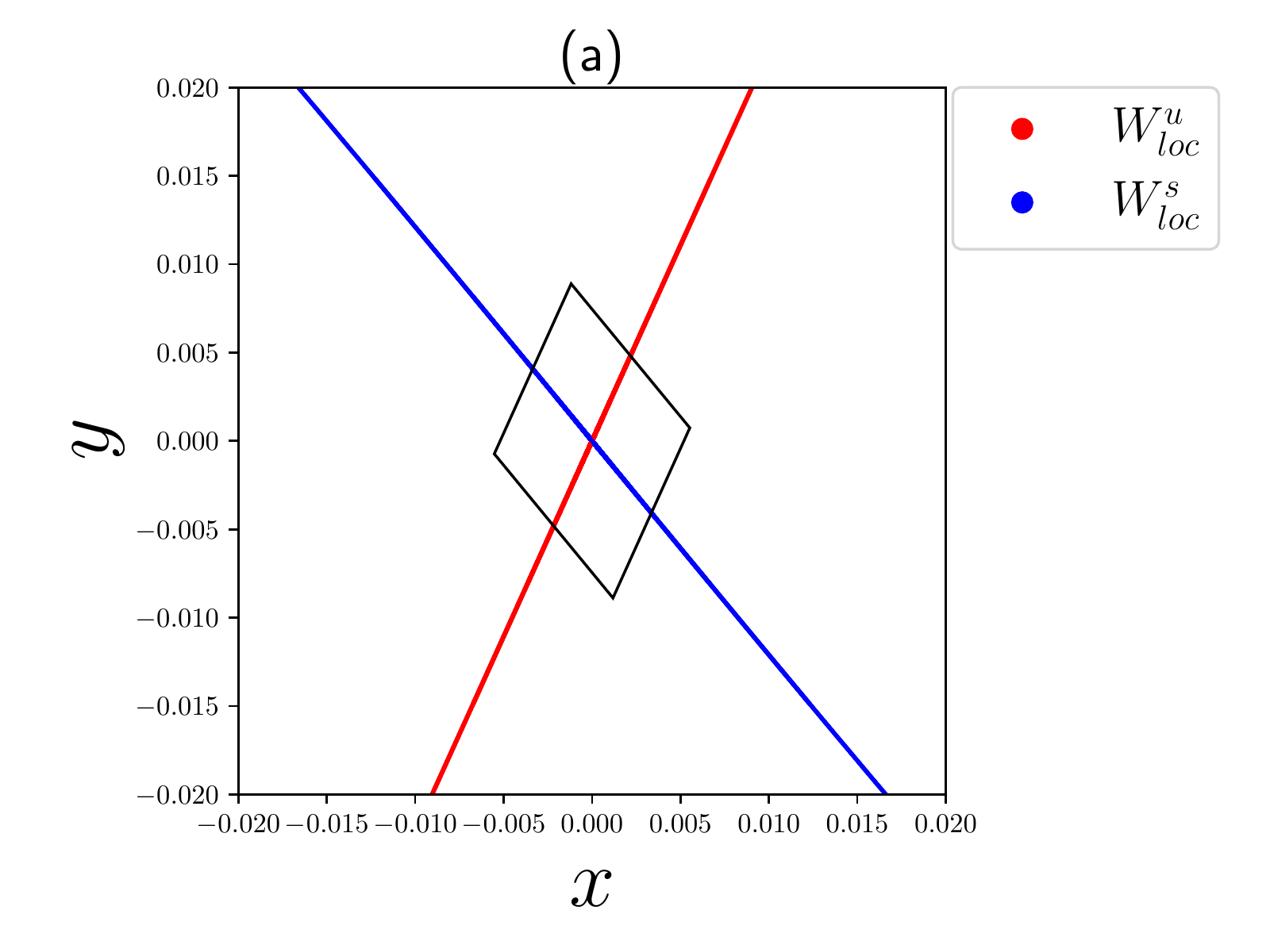}
\includegraphics[width = 7.0cm,bb= 0 0 461 346]{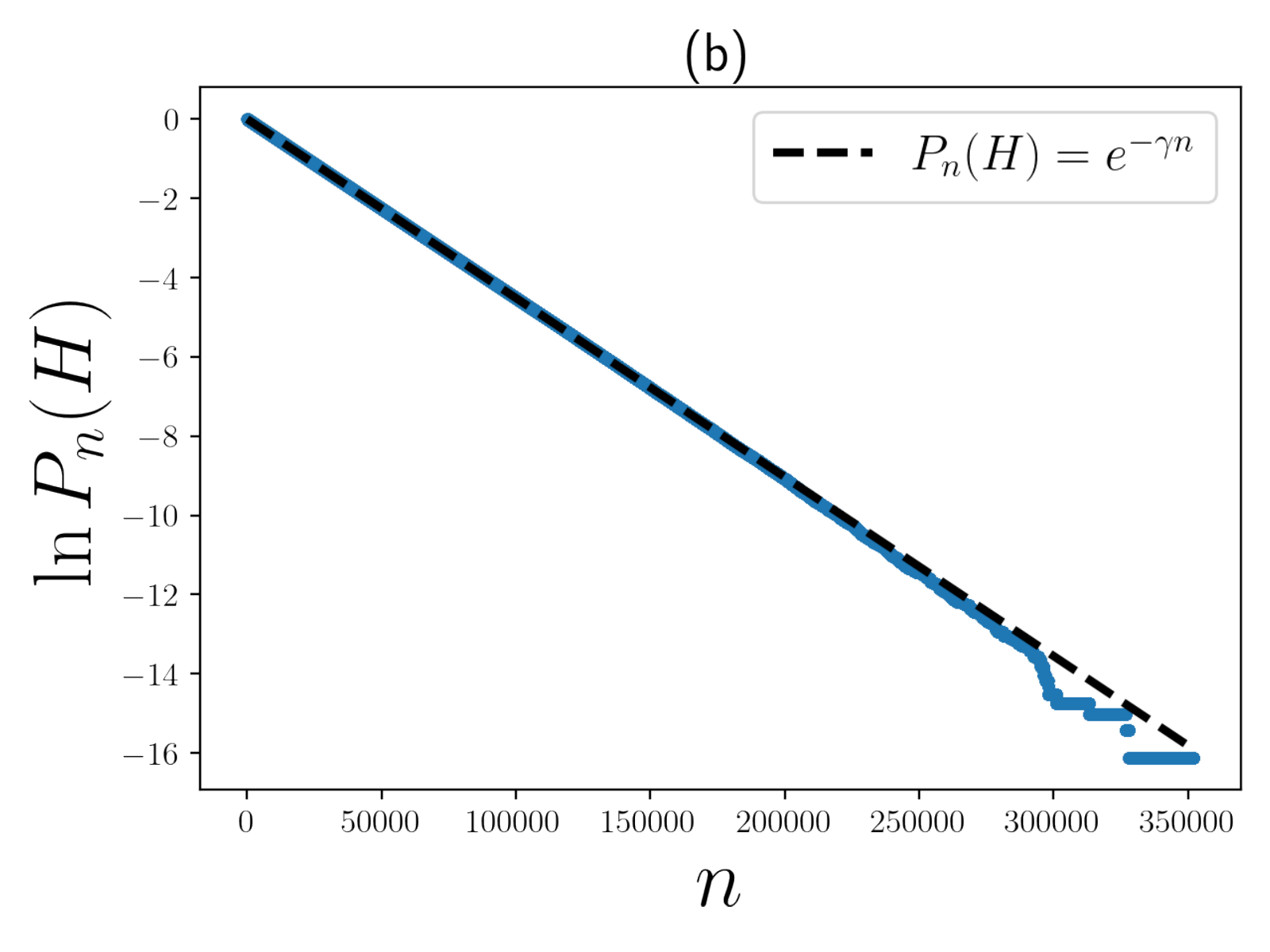}
\end{minipage}
\caption{\label{fig:fig1}
(a) The rhombus-shaped hole around the fixed point $(0, 0)$ of the perturbed cat map. The red and blue curves represent the local stable and unstable manifolds at the fixed point, respectively. (b) Survival probability $P_n(H)$ for the perturbed cat map with $\epsilon=0.1$, $\nu=2$ and $m(H)=10^{-4}$. $10^{7}$ orbits are used in evaluating $P_n(H)$. 
}
\end{center}
\end{figure}
%%%%%%%%%%%%%%%%%%%%%%%%%%%%%%%%%%%%%%%%
%%%%%%%%%%%%%%%%%%%%%%%%%%%%%%%%%%%%%%%%
%%%%%%%%%%%%%%%%%%%%%%%%%%%%%%%%%%%%%%%%

For comparison, we numerically obtain the escape rate in the following way. First, set a sufficiently small rhombus-shaped hole $H$ in the torus $\mathbb{T}^{2}$ and launch the orbits, whose initial points are uniformly distributed over the region $\mathbb{T}^{2} \backslash H$. 
We then count the $N_n$ orbits, which remain inside the region $\mathbb{T}^{2} \backslash H$ until a time step $n$. 
This yields the survival probability $P_{n}(H)=N_{n}/N_{0}$. 
We calculate the escape rate $\gamma$ by fitting the function of the form $e^{-\gamma n}$ to the survival probability $P_{n}(H)$. 
As shown in Fig.~\ref{fig:fig1}(b), the exponential decaying behavior of the survival probability 
is well achieved, which allows for evaluating the escape rate accurately.

\subsection{Derivation of equation~(\ref{eq:escape_rate})}
\label{subsec:derivation}

The escape rate $\gamma$ is written in terms of the survival 
probability $P_n(H)$ as 
\begin{eqnarray}
\label{eq:transformation of gamma}
\gamma(H)&=&-\lim_{n\rightarrow\infty}\frac{1}{n}\ln P_{n}(H) \nonumber \\
&=&-\lim_{n\rightarrow\infty}\frac{1}{n}\ln(1-m(\Xi_{n}(H))), 
\end{eqnarray}
where 
$\Xi_{n}(H):=\bigcup_{i=0}^{n}F^{-i}(H)$. 
For sufficiently small $m(\Xi_{n}(H))$, we may evaluate as 
\begin{equation}
\label{eq:Taylor series of logarithm function}
\ln(1-m(\Xi_{n}(H)))\simeq-m(\Xi_{n}(H)), 
\end{equation}
which leads to 
\begin{equation}
\label{eq:Taylor series of gamma}
\gamma(H)\simeq\lim_{n\rightarrow\infty}\frac{m(\Xi_{n}(H))}{n}. 
\end{equation}

To proceed, we introduce the inverse map $G:=F^{-1}$, 
\begin{equation}
\label{eq:inverse perturbed cat map}
G:
\left(
\begin{array}{c}
\displaystyle 
X \\
\vspace{1mm}
Y
\end{array}
\right)
\mapsto
\left(
\begin{array}{c}
\displaystyle 
2X-Y+\frac{\epsilon}{\nu}\sin(2\pi\nu(-X+Y)) \\
\vspace{1mm}
-X+Y
\end{array}
\right)
~~
\mathrm{mod}\ 1. 
\end{equation}

%%%%%%%%%%%%%%%%%%%%%%%%%%%%%%%%%%%%%%%%%
%%%%%%%%%%%%%%%%       Fig. A1        %%%%%%%%%%%%%%%
%%%%%%%%%%%%%%%%%%%%%%%%%%%%%%%%%%%%%%%%%
\begin{figure}
\begin{center}
\includegraphics[width = 7cm,bb= 0 0 461 346]{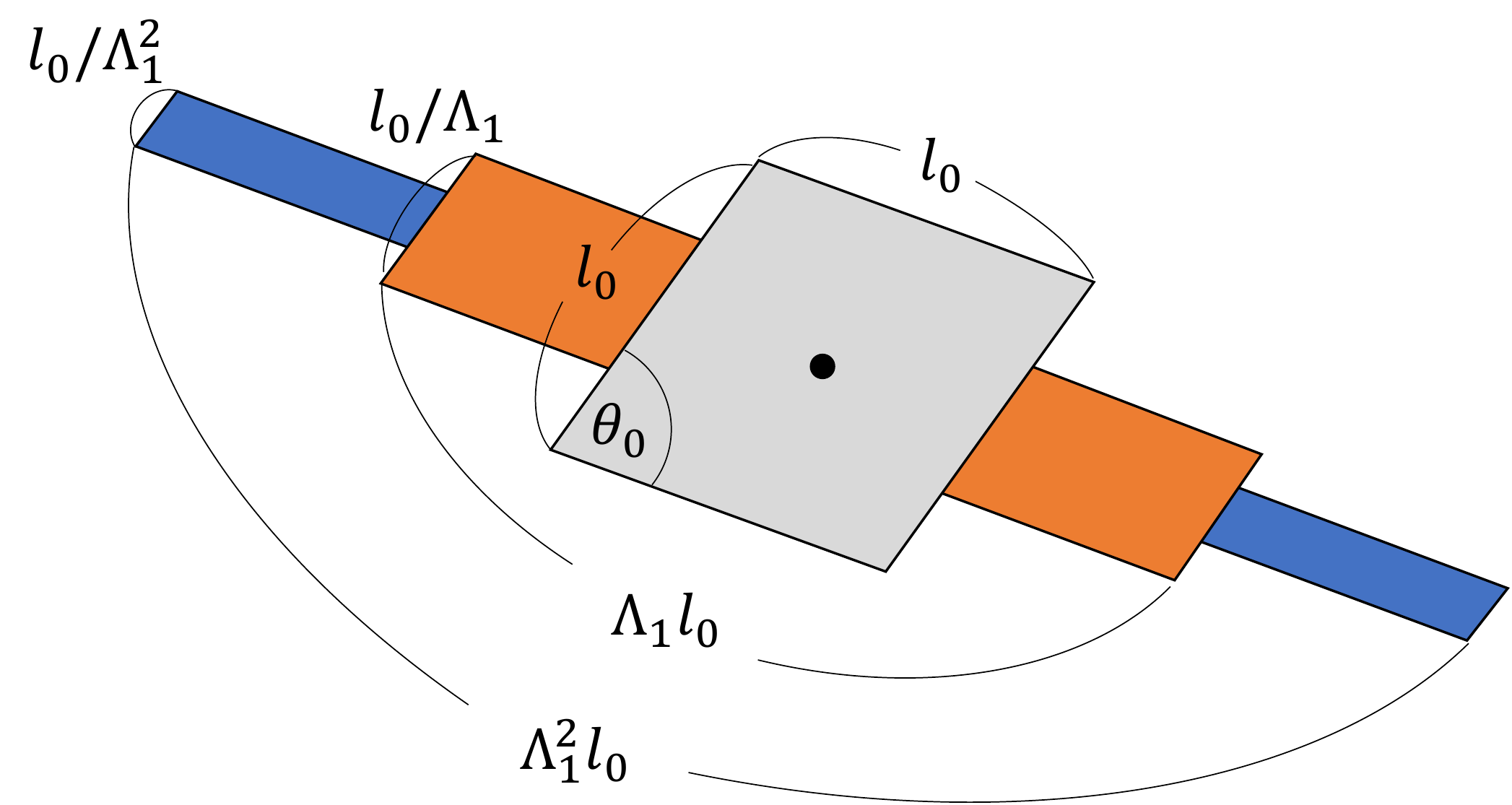}
\end{center}
\caption{
Rhombus-shaped hole $H=\Theta_{0}$ (gray), and its backward iterates 
$\Theta_{1}$ (orange) and $\Theta_{2}$ (blue) for the perturbed cat map.
%Rhombus-shaped hole $H=\Theta_{0}$, and 
%its backward iterates $\Theta_{1}$ (orange), $\Theta_{2}$ (blue), 
%$\Theta_{3}$ (green), and $\Theta_{4}$ (yellow)
% for the perturbed cat map. 
}
\label{fig:app1}
\end{figure}
%%%%%%%%%%%%%%%%%%%%%%%%%%%%%%%%%%%%%%%%%
%%%%%%%%%%%%%%%%%%%%%%%%%%%%%%%%%%%%%%%%%
%%%%%%%%%%%%%%%%%%%%%%%%%%%%%%%%%%%%%%%%%

%%%%%%%%%%%%%%%%%%%%%%%%%%%%%%%%%%%%%%%%%%
%%%%%%%%%%%%%%%%%%%%%%%%%%%%%%%%%%%%%%%%%%
\subsubsection{Fixed-point case.}
\label{appsec:fixed_point}

First, we consider the case where a fixed point $x_0$ is located at the center of the hole $H$. The hole $H$ is rhombus-shaped with sides of length $l_{0} (\ll1)$. 
One of the sides is parallel to $V^{s}$. The other is parallel to $V^{u}$, where $V^{s}$ is the eigenvector of the Jacobian $DG(x_0)$ in the stable direction and $V^{u}$ in the unstable direction, respectively. 
Note that $\theta_0$ is the angle formed by $V^{s}_{0}$ and $V^{u}_{0}$. 
Figure~\ref{fig:app1} illustrates the set $\Theta_{n}(H)$ with $0 \le n \le 2$, where 
\begin{equation}
\nonumber
\Theta_{n}(H) :=\{x\in M:F^{n}(x)\in H,  
F^{j}(x)\notin H ~{\rm for}~0 \le j \le n-1\}. 
\end{equation}
According to this definition, $\Theta_{0}(H)=H$.
We show that 
\begin{eqnarray}
\label{eq:fixed m(O0)}
m(\Xi_{0}(H))&=&l^{2}_{0}\sin\theta_{0} \nonumber \\
&=&m(H), \\
\label{eq:fixed m(O1)}
m(\Xi_{1}(H))&=&m(\Xi_{0}(H))+m(\Theta_{1}(H))\nonumber \\
&=&m(H)+m(G(H)\setminus H)\nonumber \\
&=&m(H)+l_{0}^{2}\sin\theta_{0}-l_{0}\cdot\frac{l_{0}}{\Lambda_{\rm 1}}\sin\theta_{0}\nonumber \\
%&=m(H)+l_{0}^{2}\sin\theta_{0}\left(1-\frac{1}{\Lambda_{\rm 1}}\right)\notag\\
&=&m(H)+m(H)\left(1-\frac{1}{\Lambda_{\rm 1}}\right), \\
\label{eq:fixed m(O2)}
m(\Xi_{2}(H))&=&m(\Xi_{1}(H))+m(\Theta_{2}(H))\nonumber \\
&=&m(\Xi_{1}(H))+l_{0}^{2}\sin\theta_{0} \nonumber
-\Lambda_{\rm 1}l_{0}\cdot\frac{l_{0}}{\Lambda^{2}_{\rm 1}}\sin\theta_{0}\nonumber \\
&=&m(\Xi_{1}(H))+l_{0}^{2}\sin\theta_{0}\left(1-\frac{1}{\Lambda_{\rm 1}}\right)\nonumber \\
&=&m(H)+2m(H)\left(1-\frac{1}{\Lambda_{\rm 1}}\right), 
\end{eqnarray}
where $\Lambda_{\rm 1}$ denotes the instability of the fixed point. 
For sufficiently small $l$, a similar calculation yields 
\begin{equation}
\label{eq:fixed m(On) of perturbed cat map}
m(\Xi_{n}(H))=m(H)+nm(H)\left(1-\frac{1}{\Lambda_1}\right). 
\end{equation}
Using (\ref{eq:Taylor series of gamma}), we find the expression for the escape rate: 
\begin{equation}
\label{eq:fixed gamma of perturbed cat map}
\gamma(H)\simeq m(H)\left(1-\frac{1}{\Lambda_1}\right). 
\end{equation}

%%%%%%%%%%%%%%%%%%%%%%%%%%%%%%%%%%%%%%%%%
%%%%%%%%%%%%%%%%       Fig. A2        %%%%%%%%%%%%%%%
%%%%%%%%%%%%%%%%%%%%%%%%%%%%%%%%%%%%%%%%%
\begin{figure*}
\vspace{30mm}
\begin{center}
\hspace{-70mm}
\includegraphics[width = 5.0cm,bb= 0 0 461 346]{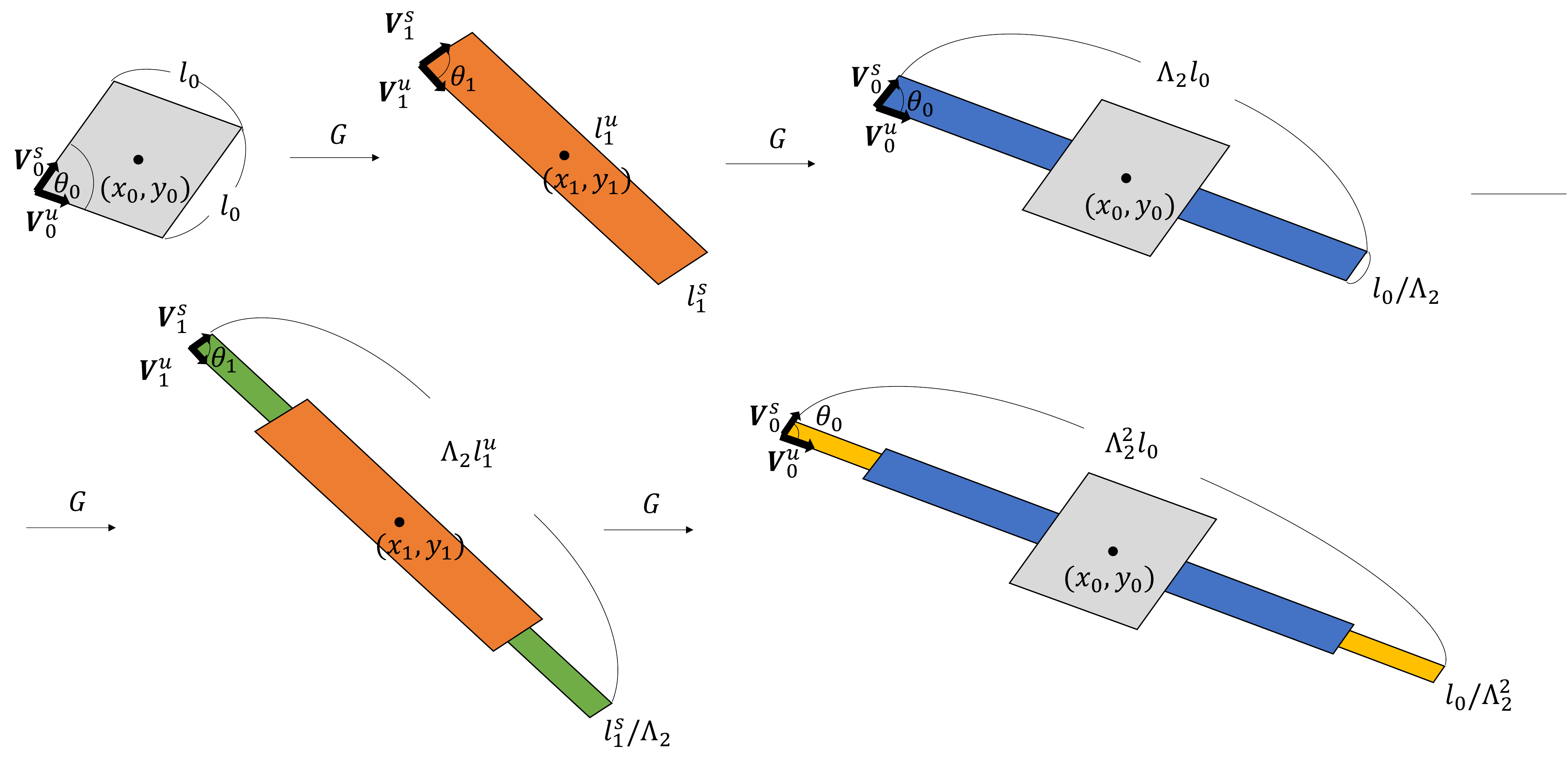}
\end{center}
\caption{
Rhombus-shaped hole $H=\Theta_{0}$ (gray), and its backward iterates $\Theta_{1}$ (orange), $\Theta_{2}$ (blue), $\Theta_{3}$ (green), and $\Theta_{4}$ (yellow) for the perturbed cat map. 
}
\label{fig:app2}
\end{figure*}
%%%%%%%%%%%%%%%%%%%%%%%%%%%%%%%%%%%%%%%%%
%%%%%%%%%%%%%%%%%%%%%%%%%%%%%%%%%%%%%%%%%
%%%%%%%%%%%%%%%%%%%%%%%%%%%%%%%%%%%%%%%%%

%%%%%%%%%%%%%%%%%%%%%%%%%%%%%%%%%%%%%%%%%%
%%%%%%%%%%%%%%%%%%%%%%%%%%%%%%%%%%%%%%%%%%
\subsubsection{Periodic-orbit case.}
\label{appsec:periodic_orbit}

We next consider a more generic case: the hole $H$ contains a point belonging to a periodic orbit with period $p >1$. For simplicity, we take the case with $p=2$. The same argument applies to higher periods as well. 

Suppose that $\{x_i=(X_{i}, Y_{i})\}_{i=0}^{1}$ is a period-2 periodic orbit and the point $x_0=(X_{0}, Y_{0})$ is contained in the rhombus-shaped hole $H$ whose side length is $l_{0} (\ll1)$. The eigenvectors $V^{s}_{0}$ and $V^{u}_{0}$ are introduced as above (see Fig.~\ref{fig:app2}). 

We next introduce the following two vectors:
\begin{eqnarray}
V_{1}^{s}&:=DG(x_0)V_{0}^{s}, \\
V_{1}^{u}&:=DG(x_0)V_{0}^{u}. 
\end{eqnarray}
Let the length of the two sides of $G(H)$ be $l_{1}^{u}$ $l_{1}^{s}$, respectively. 
Here $l_{1}^{u}>l_{1}^{s}$ is assumed. One can show that $V_{1}^{s}$ and $V_{1}^{u}$ are the eigenvalues of $DG^{2}(x_1)$. 
Note that $\theta_{i}~(i=1,2)$ denotes the angle formed by $V_{i}^{s}$ and $V_{i}^{u}$. 
Figure~\ref{fig:app2} illustrates how the set $\Xi_{n}(H)$ evolves in time. 
Since the periodic orbit instability $\Lambda_{2}$ is common for the two points $x_i~(i=1,2)$, 
it is straightforward to show that \begin{eqnarray}
\label{eq:period 2 m(O0) of perturbed cat map}
m(\Xi_{0}(H))&=&l^{2}_{0}\sin\theta_{0} \nonumber \\
&=&m(H),  \\
\label{eq:period 2 m(O1) of perturbed cat map}
m(\Xi_{1}(H))&=&m(\Xi_{0}(H))+m(\Theta_{1}(H))\nonumber \\
&=&m(H)+l_{1}^{s}l_{1}^{u}\sin\theta_{1}\nonumber \\
&=&2m(H), \\
\label{eq:period 2 m(O2) of perturbed cat map}
m(\Xi_{2}(H))&=&m(\Xi_{1}(H))+m(\Theta_{2}(H))\nonumber \\
&=&2m(H)+l_{0}^{2}\sin\theta_{0}-l_{0}\cdot\frac{l_{0}}{\Lambda_{\rm 2}}\sin\theta_{0}\nonumber  \\
%&=2m(H)+l_{0}^{2}\sin\theta_{0}\left(1-\frac{1}{\Lambda_{\rm 2}}\right)\notag\\
&=&2m(H)+m(H)\left(1-\frac{1}{\Lambda_{\rm 2}}\right), \\
\label{eq:period 2 m(O3) of perturbed cat map}
m(\Xi_{3}(H))&=&m(\Xi_{2}(H))+m(\Theta_{3}(H))\nonumber \\
&=&m(\Xi_{2}(H))+l_{1}^{s}l_{1}^{u}\sin\theta_{1}-l_{1}^{u}\cdot\frac{l_{1}^{s}}{\Lambda_{\rm 2}}\sin\theta_{1}\nonumber \\
%&=m(\Xi_{2}(H))+l_{1}^{s}l_{1}^{u}\sin\theta_{1}\left(1-\frac{1}{\Lambda_{\rm 2}}\right)\notag\\
&=&2m(H)+2m(H)\left(1-\frac{1}{\Lambda_{\rm 2}}\right).
\end{eqnarray}
Here we have used the relation $l_{1}^{s}l_{1}^{u}\sin\theta_{1}=m(G(H))=m(H)$, 
which holds because the map is area-preserving. 
For sufficiently small $l_{0}$, a similar calculation yields 
\begin{eqnarray}
\label{eq:period 2 m(On) of perturbed cat map}
m(\Xi_{n}(H))&=2m(H) \nonumber
+(n-1)m(H)\left(1-\frac{1}{\Lambda_{\rm 2}}\right).
\end{eqnarray}
Using (\ref{eq:Taylor series of gamma}), we provide an expression for 
the escape rate as
\begin{equation}
\label{eq:period 2 gamma of perturbed cat map}
\gamma(H)\simeq m(H)\left(1-\frac{1}{\Lambda_{\rm 2}}\right). 
\end{equation}

In a similar way, in the case where the hole $H$ contains 
a  point $x_0=(X_{0}, Y_{0})$ of a periodic orbit $\{x_i=(X_{i}, Y_{i})\}_{i=0}^{p-1}$, 
we introduce 
\begin{eqnarray}
&V_{i}^{s}:=DG(x_{i-1})\cdots DG(x_0)V_{0}^{s}, \\
&V_{i}^{u}:=DG(x_{i-1})\cdots DG(x_0)V_{0}^{u}.
\end{eqnarray}
Since these vectors are the eigenvectors of $DG^{p}(x_i)$, 
we find 
\begin{equation}
\label{eq:period p m(On) of perturbed cat map}
m(\Xi_{n}(H)) =p\,m(H) +(n-(p-1)) 
 m(H)\left(1-\frac{1}{\Lambda_{p}}\right), 
\end{equation}
which leads to 
\begin{equation}
\label{eq:period p gamma of perturbed cat map}
\gamma(H)\simeq m(H)\left(1-\frac{1}{\Lambda_{p}}\right). 
\end{equation}

\subsection{Numerical verification of the escape rate formula}
\label{sec:numerical_verification}

\begin{figure*}[htb!]
\begin{center}
\begin{minipage}{1\hsize}
\includegraphics[width = 7.0cm,bb= 0 0 461 346]{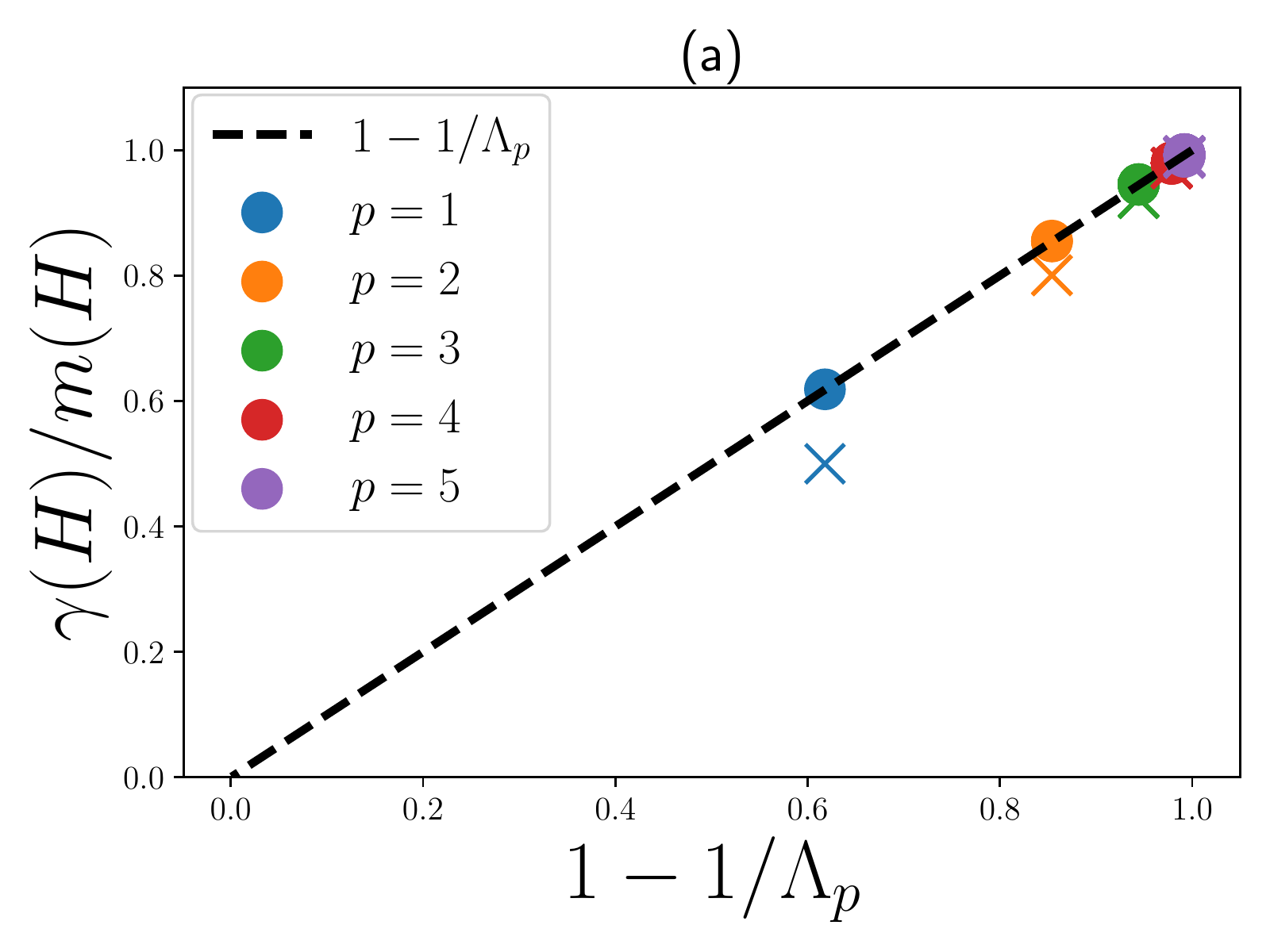}
%\hspace{-20mm}
\includegraphics[width = 7.0cm,bb= 0 0 461 346]{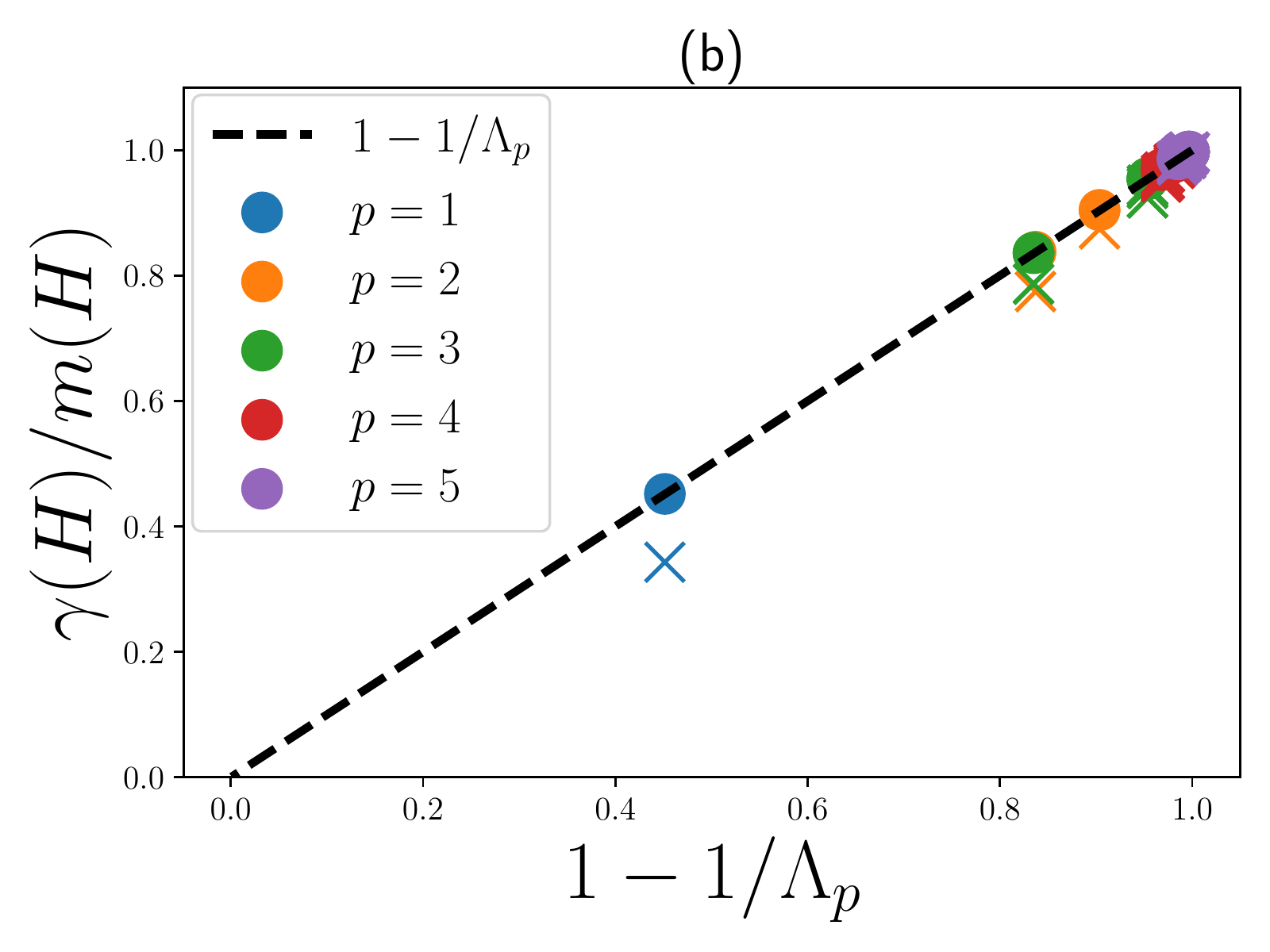}
\includegraphics[width = 7.0cm,bb= 0 0 461 346]{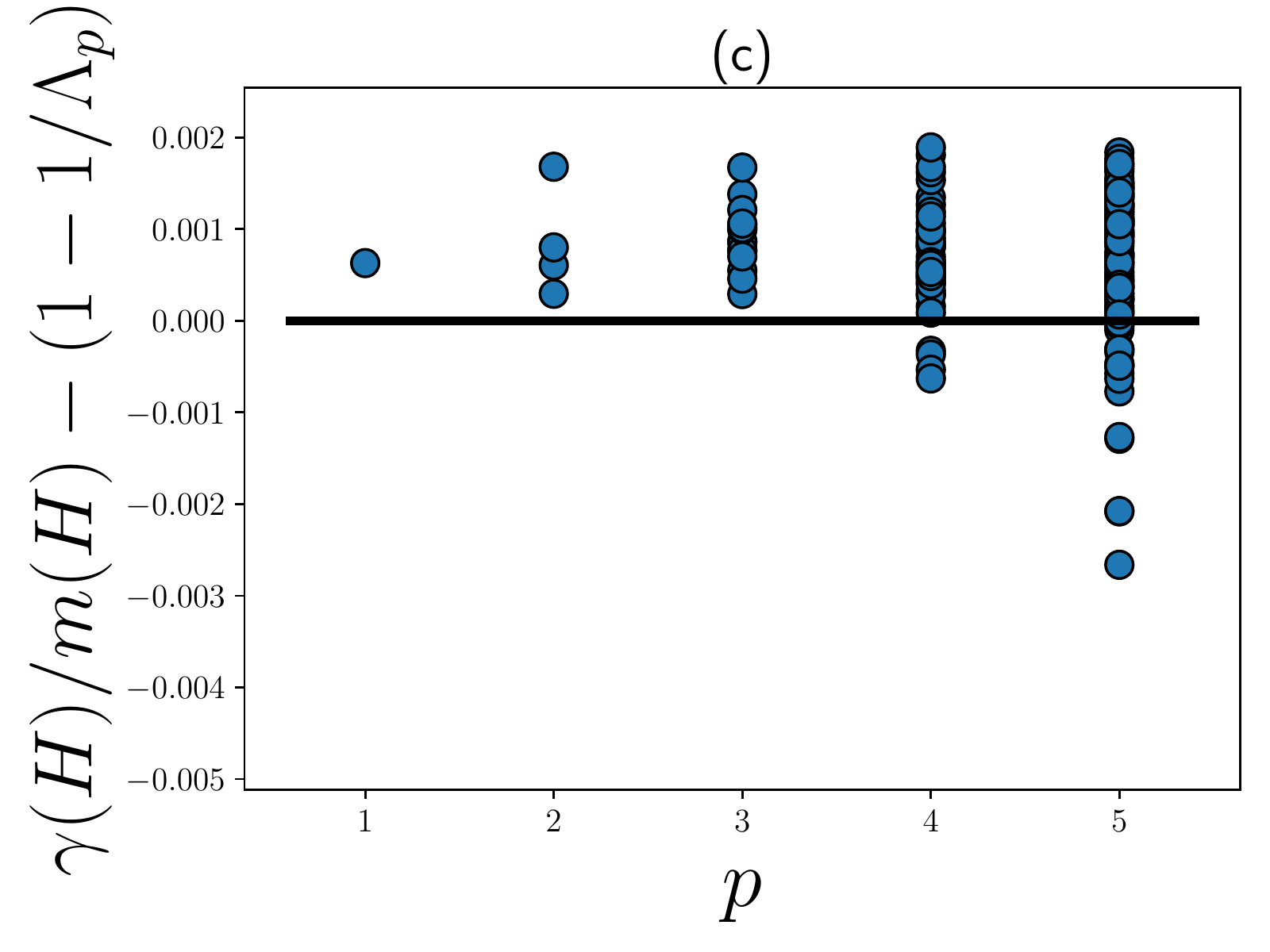}
%\hspace{1.5mm}
\includegraphics[width = 7.0cm,bb= 0 0 461 346]{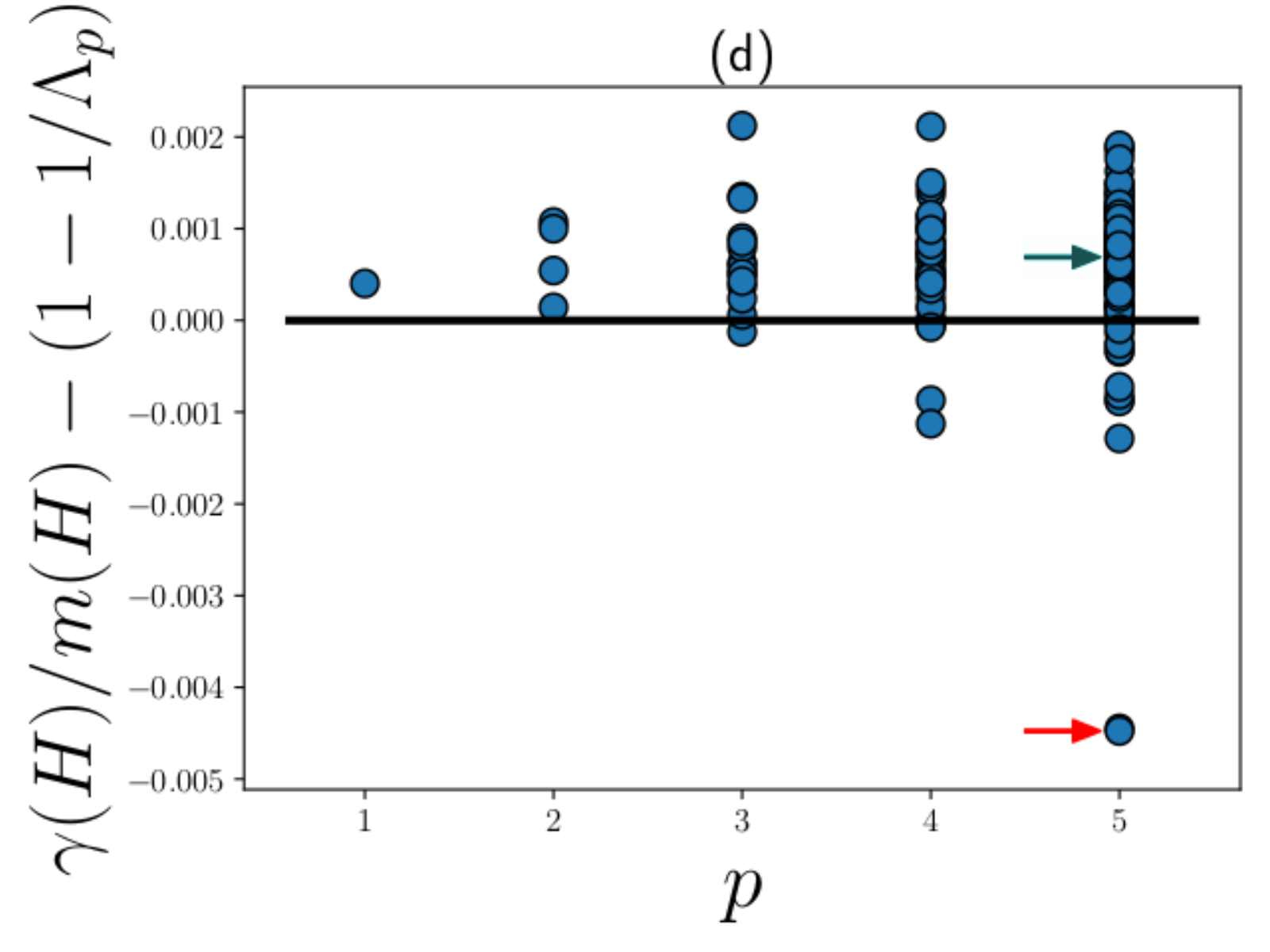}
\end{minipage}
\caption{\label{fig:fig2}
Normalized escape rate for the (a) unperturbed and (b) perturbed cat map with $\epsilon=0.1$, $\nu=2$ and $m(H)=10^{-4}$, plotted as a function of $1 - 1/\Lambda_{p}$. 
The colored circles represent the results for the rhombus-shaped hole whose sides are respectively oriented in local stable and unstable directions at the center of the hole. 
The crosses denote the results for the square-shaped hole whose sides are not oriented in local stable and unstable directions. 
$p$ denotes the period of the periodic orbit located at the center of the hole $H$. 
There are multiple periodic orbits per period, but we cannot distinguish escape rates that result from placing the hole on different cycles of the same period,  up to the resolution of the figures (a) and (b). This is achieved instead in   
(c) and (d) for a rhombus-shaped hole, by  plotting the deviation of the numerical results from the formula (\ref{eq:escape_rate}),
each plot referring to the graph above it.   
The cases indicated by the red arrow (two overlapping points) and green arrow in (d) are closely examined in Fig.~\ref{fig:fig3} and
Fig.~\ref{fig:fig4}, respectively. 
%Note that almost indistinguishable two points are plotted there.
}
\end{center}
\end{figure*}

In this subsection, we numerically check the validity of the formula (\ref{eq:escape_rate}). 
Figure~\ref{fig:fig2} compares the formula (\ref{eq:escape_rate}) with the numerical results 
obtained by applying the method explained above. 
It is apparent that Eq.~(\ref{eq:escape_rate}) well predicts the numerical results, with rare exceptions illustrated in Figs.~\ref{fig:fig2}(c) and (d). 
It is noted that the number of periodic orbits increases exponentially with the period $p$, consistently with the positive topological entropy. 

Before investigating the origin of the discrepancies, let us provide the result obtained when replacing the rhombus-shaped box with the square whose sides are taken simply in the horizontal and vertical direction, respectively. 
As displayed in Fig.~\ref{fig:fig2}(a) and (b), the numerically obtained escape rate into the square hole (colored crosses) significantly deviates from the theoretical prediction even for a small area of the leak. 
This implies that the hole shape should be carefully chosen when one seeks the validity of a rigorous formula, as pointed out in \cite{paar2000bursts}. Periodic orbits of different periods may occasionally have similar instabilities, which lead to similar escape rates [Fig.~\ref{fig:fig2}(b)], consistently with Eq.~(\ref{eq:escape_rate}).
%in the FDL hyperbolic system \cite{bunimovich2012fair}. 

%%%%%%%%%%%%%%%%%%%%%%%%%%%%%%%%%%%%%%%%
%%%%%%%%%%%%%%%       Fig. 2        %%%%%%%%%%%%%%%%
%%%%%%%%%%%%%%%%%%%%%%%%%%%%%%%%%%%%%%%%
%%%%%%%%%%%%%%%%%%%%%%%%%%%%%%%%%%%%%%%%
%%%%%%%%%%%%%%%%%%%%%%%%%%%%%%%%%%%%%%%%
%%%%%%%%%%%%%%%%%%%%%%%%%%%%%%%%%%%%%%%%

\begin{figure}
\begin{center}
\includegraphics[width = 6.5cm,bb= 0 0 461 346]{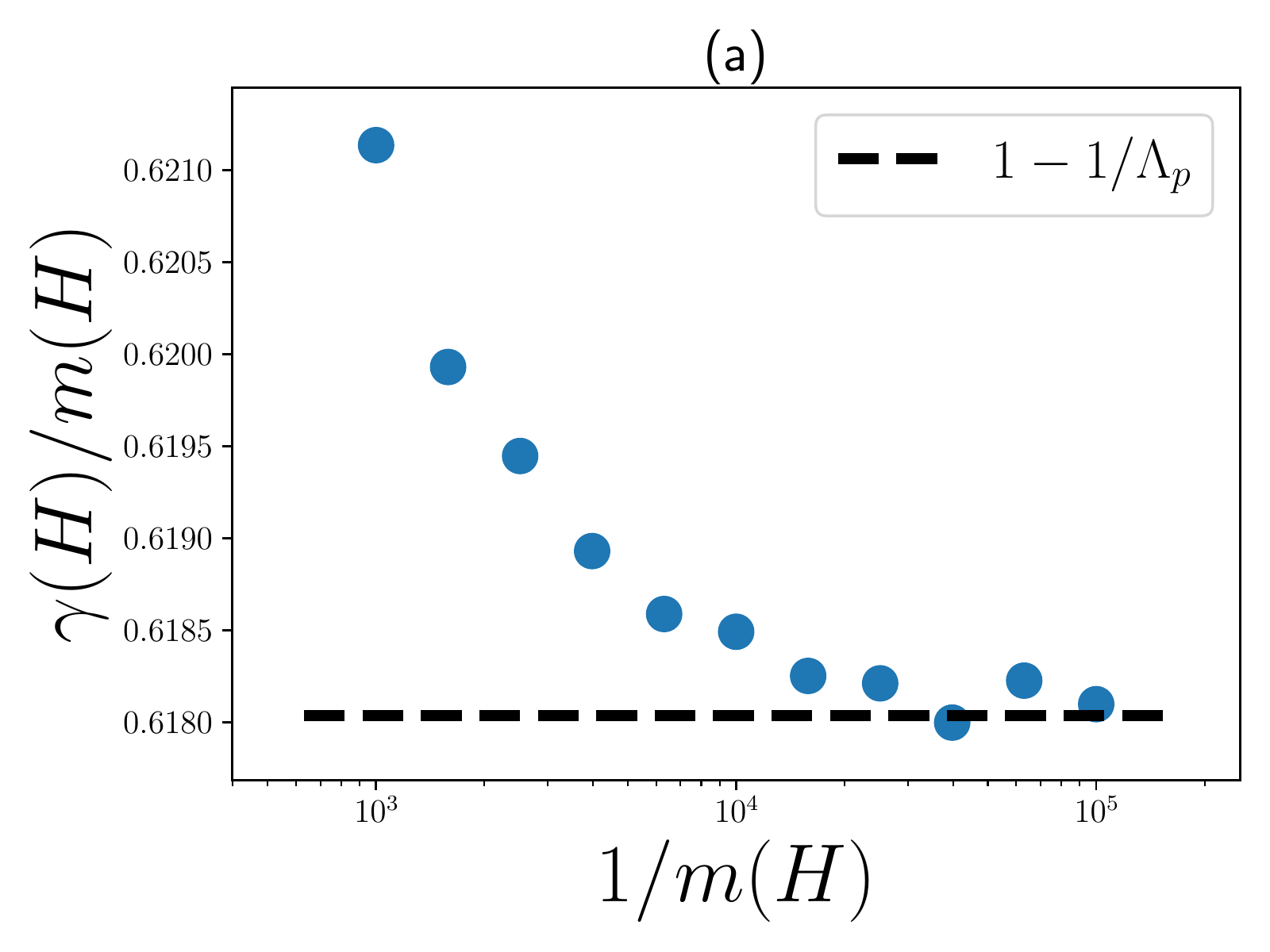}
\includegraphics[width = 6.5cm,bb= 0 0 461 346]{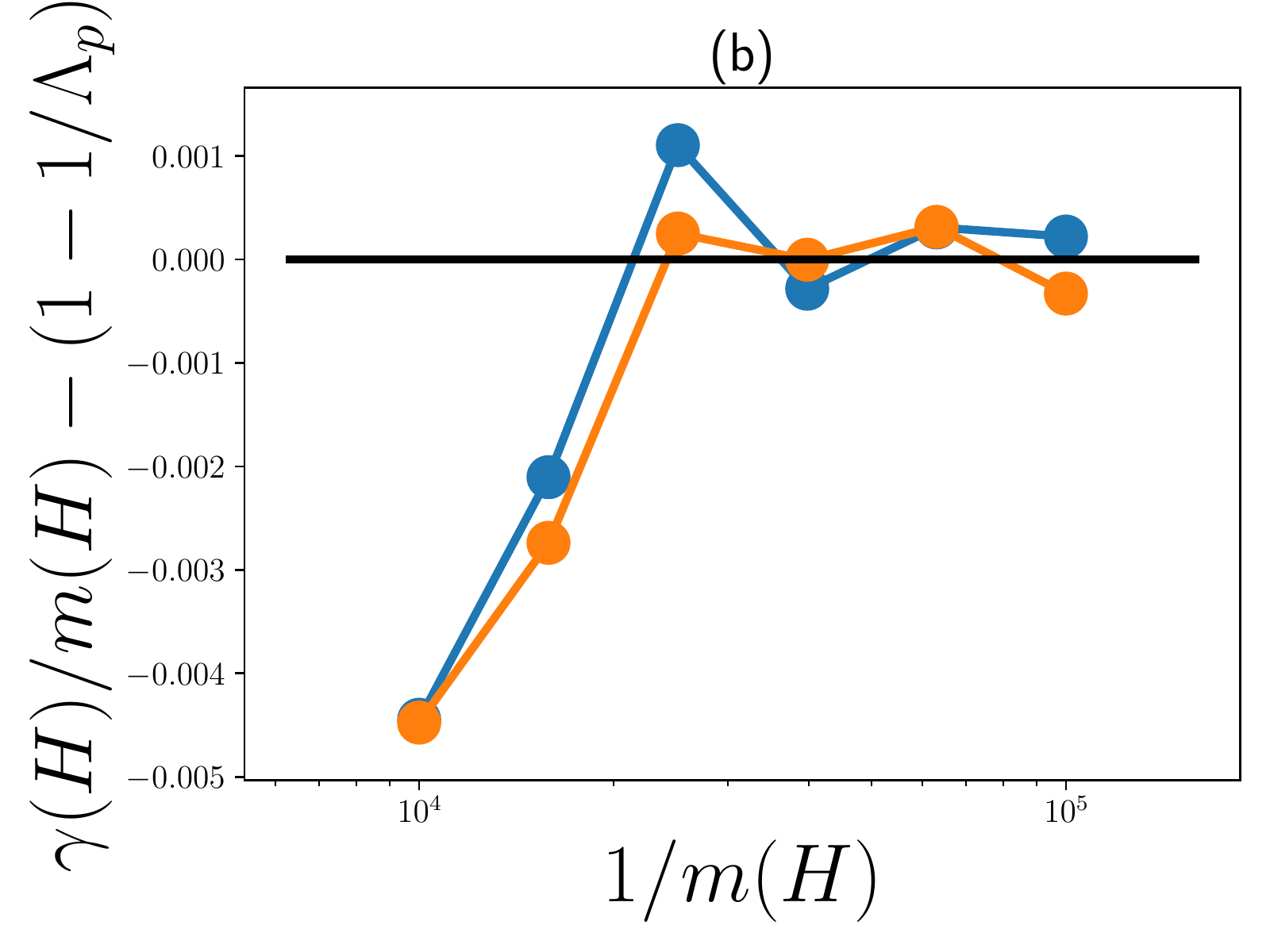}
\end{center}
\caption{
(a) Normalized escape rate as a function of the inverse $1/m(H)$ of the area of the  (rhombus-shaped)  leak 
for the cat map. 
%Inset: blow-up of the region inside the red box of the main graph. 
%is put to closely see a larger $1/m(H)$ region. 
The dashed line indicates the theoretical prediction (\ref{eq:escape_rate}). 
(b) %Normalized escape rate as a function of the inverse $1/m(H)$ of the hole size
%for the perturbed cat map for 
Deviation of the numerical results from the formula (\ref{eq:escape_rate}) for
the two period-5 cycles indicated by  the red arrow in 
Fig.~\ref{fig:fig2}(d). The parameters are chosen as $\epsilon=0.1$ and $\nu=2$. 
}
\label{fig:fig3}
\end{figure}

As for the rhombus-shaped hole, we wish to verify that the deviations observed in Figs.~\ref{fig:fig2}(c) and (d) come solely from the finiteness of the area of the leak, and thus we reduce the latter and see how the escape rate $\gamma$ behaves. 
Figure~\ref{fig:fig3}(a) plots the normalized escape rate $\gamma(H)/m(H)$ as a function of the inverse area of the leak $1/m(H)$. 
The result clearly shows that the escape rate tends to the value predicted by the formula (\ref{eq:escape_rate}) as the area of the leak decreases.

The result presented in Fig.~\ref{fig:fig3}(a) suggests that the correction to the leading-order is positive, since numerically obtained escape rates exceed the leading-order prediction in most values of $1/m(H)$. 
This is consistent with the result for the doubling map \cite{georgiou2012faster}. 

%%%%%%%%%%%%%%%%%%%%%%%%%%%%%%%%%%%%%%%%%
%%%%%%%%%%%%%%%%       Fig. 3        %%%%%%%%%%%%%%%%
%%%%%%%%%%%%%%%%%%%%%%%%%%%%%%%%%%%%%%%%%
%%%%%%%%%%%%%%%%%%%%%%%%%%%%%%%%%%%%%%%%%
%%%%%%%%%%%%%%%%%%%%%%%%%%%%%%%%%%%%%%%%%
%%%%%%%%%%%%%%%%%%%%%%%%%%%%%%%%%%%%%%%%%

\begin{figure}[ht]
\begin{center}
\begin{minipage}{1\hsize}
\hspace{15mm}
\includegraphics[width = 6.5cm,bb= 0 0 461 346]{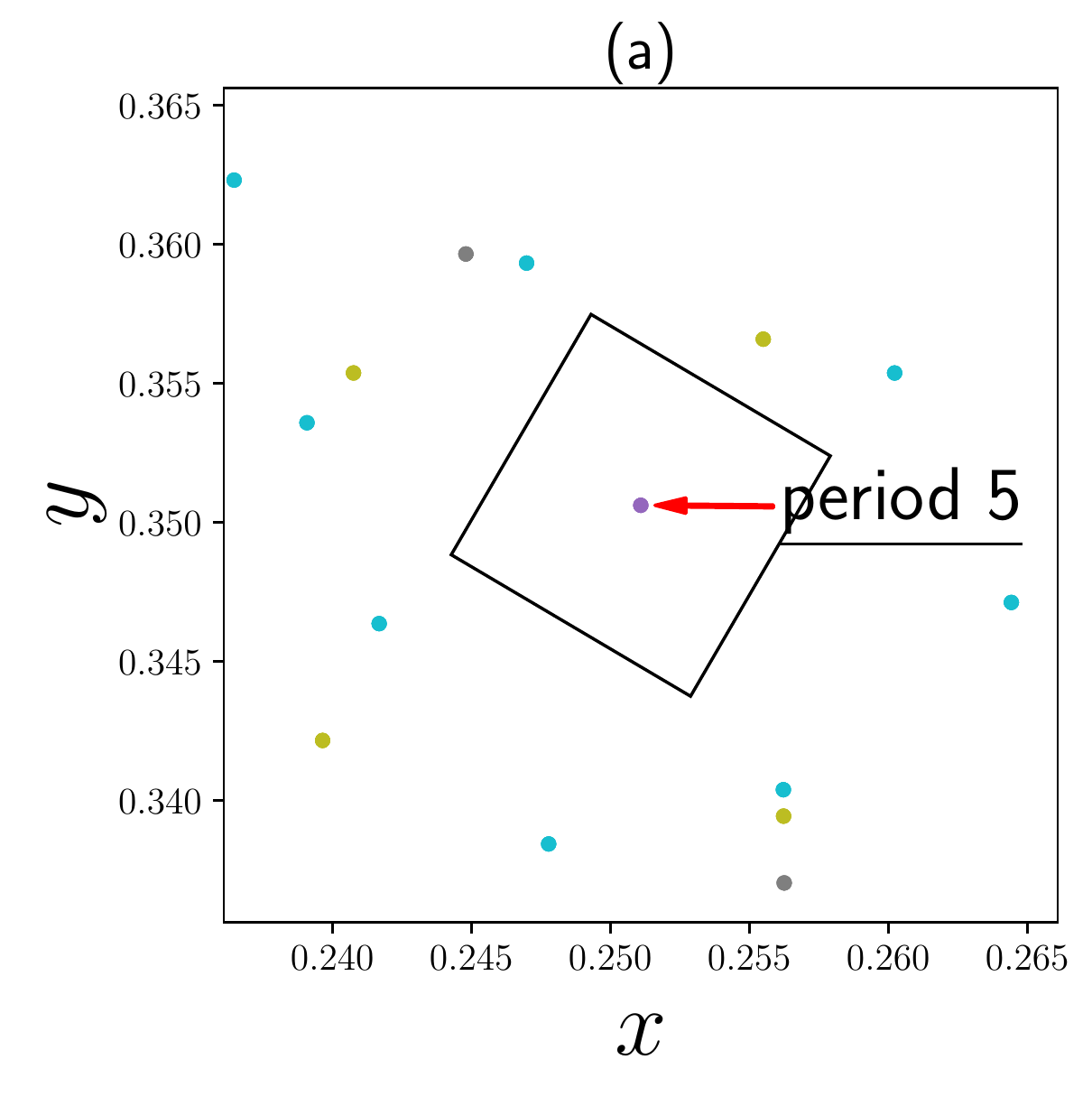}
\includegraphics[width = 6.5cm,bb= 0 0 461 346]{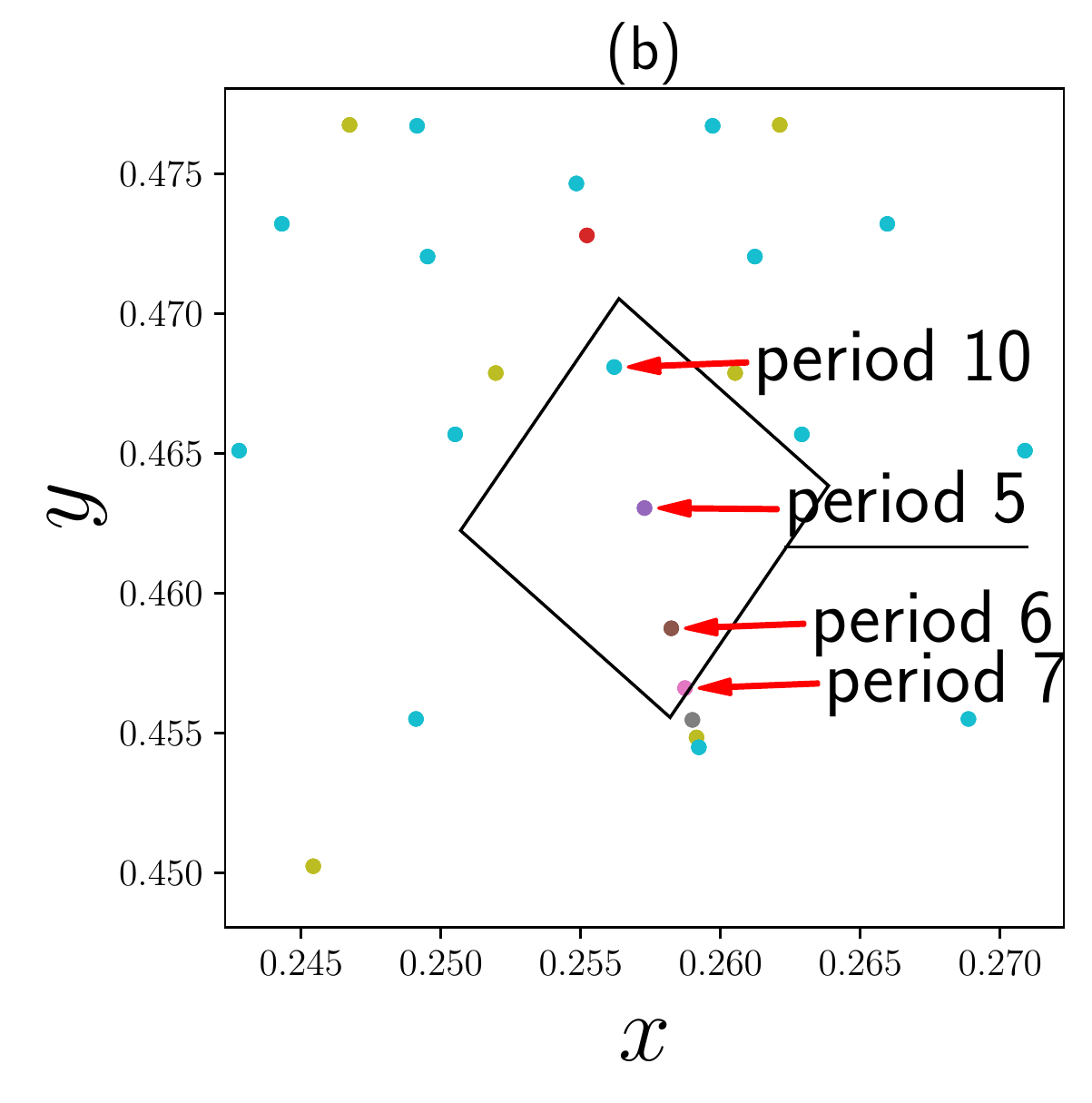}
\end{minipage}
\caption{
Periodic points of orbits of period up to $p=10$, lying inside- and in the vicinity of the hole $H$, that has a point of a period-5 orbit at its center. 
In (a) the value of the escape rate is very close to the theoretical prediction (\ref{eq:escape_rate}), shown by the green arrow in Fig.~\ref{fig:fig2}(d).
In (b) the numerically obtained escape rate is far below the theoretical prediction (\ref{eq:escape_rate}), shown by the red arrow in Fig.~\ref{fig:fig2}(d). 
}
\label{fig:fig4}
\end{center}
\end{figure}

We can also explain why some numerical values for $p=4$ and $p=5$ are smaller than the theoretical prediction. 
As seen in~subsection \ref{subsec:derivation}, the formula (\ref{eq:escape_rate}) 
has been derived assuming that only a single periodic point lies in the hole $H$ and other periodic points are not taken into account. 
Figure~\ref{fig:fig4} reveals that in the case where only the target periodic point lies in the hole $H$, the obtained escape rate does not deviate from the theoretical prediction, whereas the numerical values are 
significantly below the expected value when several periodic orbits with 
relatively short periods are contained in the hole $H$. 
The contribution of such additional periodic orbits could be avoided by reducing the area of the leak.  Figure~\ref{fig:fig3}(b) provides evidence validating that this is indeed the case: the difference between the numerical values and the theoretical prediction becomes smaller as the area of the leak $m(H)$ decreases.
The observations above partly echo with the rigorous results obtained in~\cite{bunimovich2011place} for one-dimensional maps with uniform instability, where the escape rate was found to decrease monotonically with the inverse size of the hole shrinking discretely as $2^{-N}$, $N\in\mathbb{Z}$.     
 Prospectively,
it would be worth examining the effect of higher-order corrections, which have been studied for uniformly hyperbolic systems admitting a finite Markov partition~\cite{georgiou2012faster}.

%%%%%%%%%%%%%%%%%%%%%%%%%%%%%%%%%%%%%%%%%
%%%%%%%%%%%%%%%%       Fig. 4         %%%%%%%%%%%%%%%%
%%%%%%%%%%%%%%%%%%%%%%%%%%%%%%%%%%%%%%%%%
%%%%%%%%%%%%%%%%%%%%%%%%%%%%%%%%%%%%%%%%%
%%%%%%%%%%%%%%%%%%%%%%%%%%%%%%%%%%%%%%%%%
%%%%%%%%%%%%%%%%%%%%%%%%%%%%%%%%%%%%%%%%%

\section{Escape rate for noisy systems}
\label{sec:escape_rate_with_noise}

\subsection{Numerical observation}
\label{sec:numerical_results_noise}

In the previous section, we showed that the formula (\ref{eq:escape_rate}) for the escape rate with a contribution from the shortest periodic orbit works if one takes a sufficiently small hole. 
This confirms that the escape rate is strongly affected by short periodic orbits located in the hole even though the system is uniformly hyperbolic. 
In this section, we examine the effect of noise on the escape rate. 
Instead of the map (\ref{eq:perturbed_cat_map}), we here introduce the map in the following form:
\begin{equation}
\label
{eq:perturbed_cat_map_noise}
F_{n}:
\left(
\begin{array}{c}
\displaystyle 
X
\vspace{1mm}
\\
\displaystyle 
Y
\end{array}
\right)
\mapsto
\left(
\begin{array}{c} 
\displaystyle 
X+Y-\frac{\epsilon}{\nu}\sin(2\pi\nu Y)+\omega_{n}^{(X)}
\vspace{1mm}
\\
\displaystyle 
X+2Y-\frac{\epsilon}{\nu}\sin(2\pi\nu Y)+\omega_{n}^{(Y)}
\end{array}
\right)
~~
\mathrm{mod}\ 1, 
\end{equation}
where $\omega_{n}^{(X)}$ and $\omega_{n}^{(Y)}$ denote uniform random numbers generated from the interval $[-\Omega/2, \Omega/2)$. 
The random perturbation is applied at each iteration, and $\omega_{n}^{(X)}$ and $\omega_{n}^{(Y)}$ are independent of each other. 
We call $\Omega$ the noise strength or amplitude hereafter. 

\begin{figure}
\begin{minipage}{0.5\hsize}
\begin{center}
\includegraphics[width = 7.0cm,bb= 0 0 461 346]{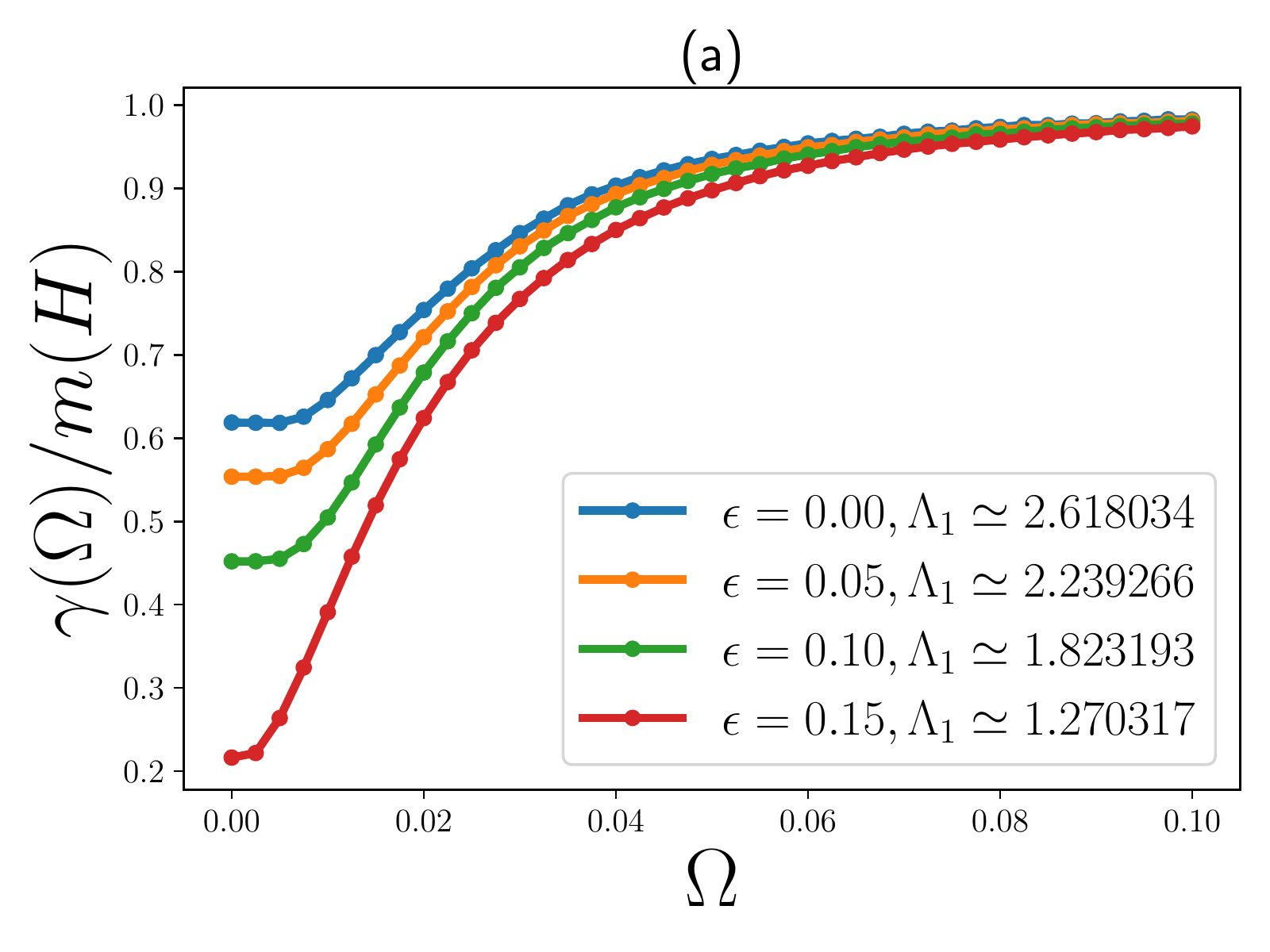}
\end{center}
\end{minipage}
\begin{minipage}{0.5\hsize}
\begin{center}
\includegraphics[width = 7.0cm,bb= 0 0 461 346]{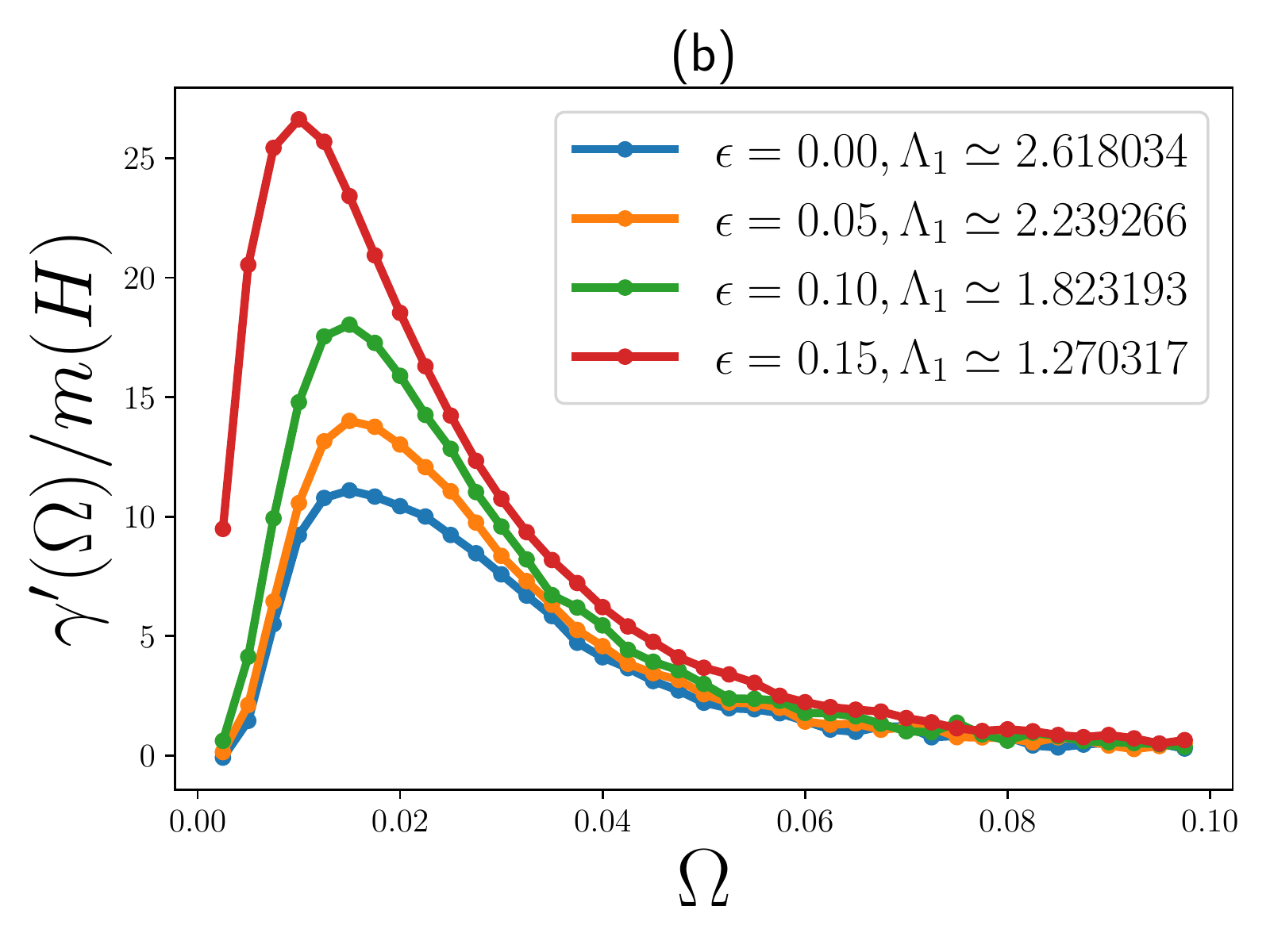}
\end{center}
\end{minipage}
\caption{
(a) Numerically computed normalized escape rate $\gamma(\Omega)/m(H)$ and (b) its derivative $\gamma' (\Omega)/m(H)$ plotted as a function of the noise strength. 
$\Lambda_1$ denotes the instability of the fixed point contained in the hole for the cat map and perturbed cat map respectively, with $\nu=2$.  The area of the leak is $m(H)=10^{-4}$. 
}
\label{fig:fig5}
\end{figure}

For simplicity, we consider a hole with the fixed point of the map at its center. 
In the perturbed cat map, the instability of the fixed point decreases with the increase of the perturbation parameter $\epsilon$; thus, the escape rate is expected to increase according to the formula (\ref{eq:escape_rate}). 
As illustrated in Fig.~\ref{fig:fig5}(a), at $\Omega=0$ the escape rate is lined up in the predicted order. 

Next, noise is added: it is first noted that the amplitudes investigated here range from very weak to strong, as it can be inferred by comparing the range of values of $\Omega^2$ on the 
abscissae of the graphs in Fig.~\ref{fig:fig5} with the selected area of the leak, $m(H)=10^{-4}$.  
As the noise strength $\Omega$ increases, as seen in Fig.~\ref{fig:fig5}(a), the escape rate overall grows and eventually saturates to the value of the area of the leak $m(H)$, when the escape becomes driven only by noise. Hence, for a sufficiently large noise strength limit, the escaping process is purely dominated by noise, and no signatures of the underlying dynamics remain. 
Here, we notice two features: 
1) a plateau appears before the escape rate grows as a function of the noise strength $\Omega$, and its length decreases %becomes smaller with an increase of 
with $\epsilon$; 
2) the response to noise, measured by the derivative $\gamma'(\Omega)$, is sharper as the value of $\epsilon$ grows (see Fig.~\ref{fig:fig5}(b)). 
These numerical results reveal that the deterministic instability of the periodic orbits contained in the hole still affects the escape rate, even in the presence of noise.

%%%%%%%%%%%%%%%%%%%%%%%%%%%%%%%%%%%%%%%%%%
%%%%%%%%%%%%%%%%%       Fig. 5        %%%%%%%%%%%%%%%%
%%%%%%%%%%%%%%%%%%%%%%%%%%%%%%%%%%%%%%%%%%
%%%%%%%%%%%%%%%%%%%%%%%%%%%%%%%%%%%%%%%%%
%%%%%%%%%%%%%%%%%%%%%%%%%%%%%%%%%%%%%%%%%
%%%%%%%%%%%%%%%%%%%%%%%%%%%%%%%%%%%%%%%%%

\subsection{Local model}
\label{sec:local_model}

\begin{figure}
\vspace{20mm}
\begin{center}
\includegraphics[width = 4.0cm,bb= 0 0 461 346]{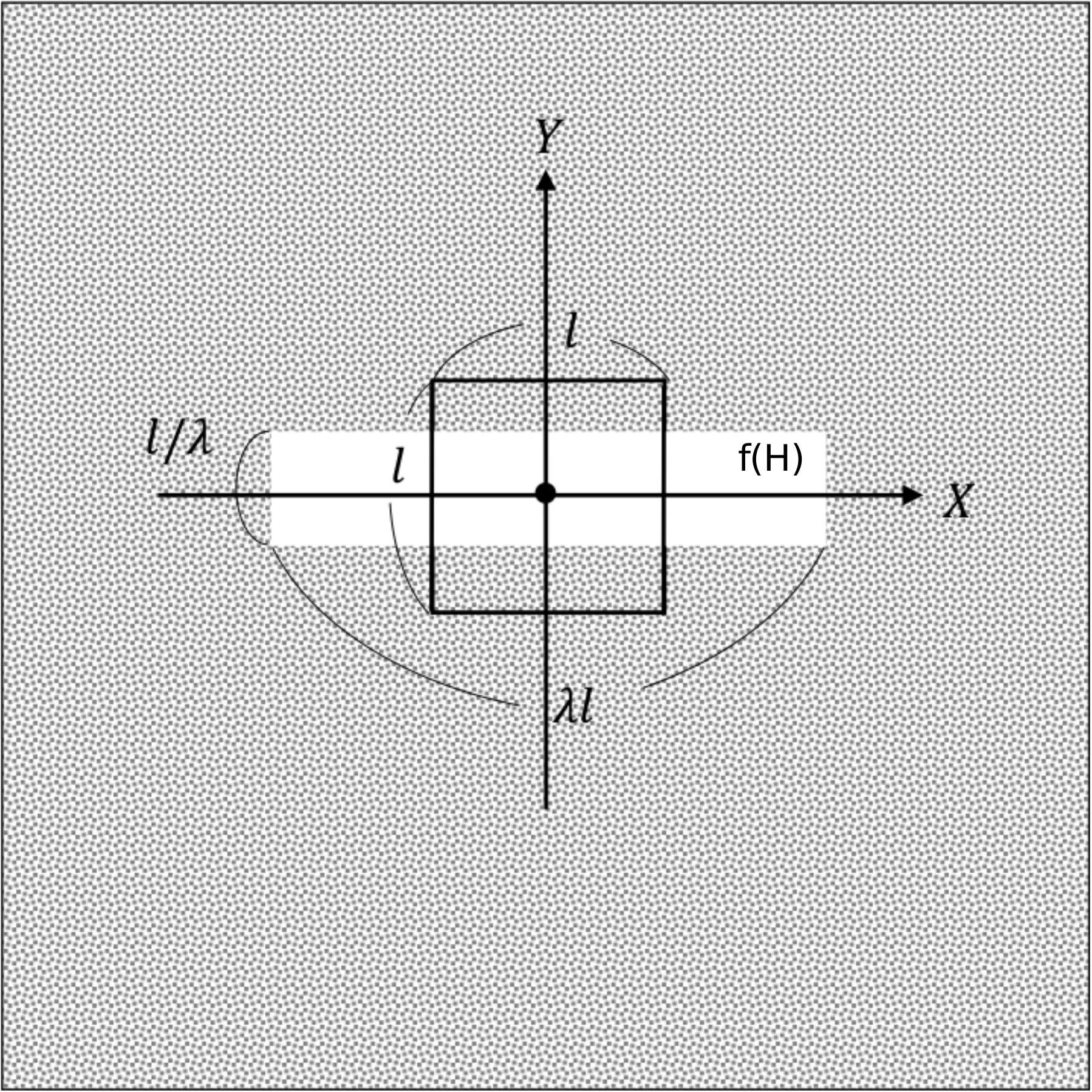}
\end{center}
\caption{
The distribution of orbits just before applying the noise $(\omega_{n}^{(X)}$, $\omega_{n}^{(Y)})$. The orbits contained in $M\setminus f(H)$ are shown in gray, and those 
contained in $f(H)$ in white. 
}
\label{fig:fig6}
\end{figure}

In order to understand the behavior of the escape rate presented in Fig.~\ref{fig:fig5}, we here introduce a local model. 
Suppose that an unstable fixed point $(X, Y) = (0,0)$ is contained in a $l \times l$ square-sharped hole $H$ whose sides are oriented in the local stable and unstable directions. 
Let $M$ be the space on which the map $f$ acts. 
We further assume that the local instability is $\lambda>1$ and the associated unstable direction is taken horizontally, and the stable direction is taken vertically (see Fig.~\ref{fig:fig6}). 

%%%%%%%%%%%%%%%%%%%%%%%%%%%%%%%%%%%%%%%%%%
%%%%%%%%%%%%%%%%%       Fig. 6         %%%%%%%%%%%%%%%%
%%%%%%%%%%%%%%%%%%%%%%%%%%%%%%%%%%%%%%%%%%
%%%%%%%%%%%%%%%%%%%%%%%%%%%%%%%%%%%%%%%%%
%%%%%%%%%%%%%%%%%%%%%%%%%%%%%%%%%%%%%%%%%

Let $\Theta_{n, \Omega}(H)$ be the set of the initial points in $M$ 
that escape from the hole $H$ exactly at the time step $n$, and $\Xi_{n, \Omega}(H)$ 
the set of points that escape at  time step $n$ or earlier. 
Here we assume that 
\begin{enumerate}
\item the orbits are uniformly distributed in the region $M \setminus H$ for $n \ge 0$. 

\item
We add noise $(\omega_{n}^{(X)}, \omega_{n}^{(Y)})$ after each iteration of the map $f$. 
\end{enumerate}
From the first assumption, the orbits before adding noise are distributed in the way illustrated in Fig.~\ref{fig:fig6}. This is an approximation, since in the deterministic system
the long time decay is governed by a nonattracting chaotic set, which is a fractal. For
any leak, there is a set of never escaping points in the complement set to the hole. In area-preserving maps
this is a chaotic saddle that possesses fractal stable and unstable manifolds, and the mentioned
conditionally invariant density sits on the unstable manifold. Noise does smudge these fractal sets,
which nontheless still constitute the skeleton of the dynamics.
 
Our task is that to evaluate $m\left(\Theta_{n, \Omega}(H)\right)$, in order to then determine the escape rate. 
In the absence of noise, we may compute the measure of the set of points that escape in one iteration by means of the 
Perron-Frobenius operator~\cite{chaosbook}, as the overlap of the iterated available phase space and the opening~\footnote{An alternative approach would be that to make $f(x)$ also depend on noise and then integrate over the ensemble~\cite{BeSz88}.}:
\begin{eqnarray}
\nonumber
\hspace{-10mm}
m\left(\Theta_{1}(H)\right) &=& \int_M \capchi_H(y)\,dy\int_M dx\, \delta(y-f(x))\capchi_{M\setminus H}(x) 
\\ &=&
\int_M dx\, \capchi_H\left(f(x)\right)\, \capchi_{M\setminus H}(x)
= m\left( f^{-1}(H) \bigcap \left[M\setminus H\right]\right)
\,,  
\label{detTheta1}
\end{eqnarray}
where 
\begin{equation}
\capchi_A(x) = \left\{
\begin{array}{cc}
1 & x\in A \\
0 & x\notin A
\end{array}
\right.
\,.
\end{equation}
%After adding noise, some orbits, which are supposed not to enter the hole under the dynamics $f$, fall into the hole, and some other orbits behave in an opposite way. 
In the presence of noise, we leave the dynamics (kernel of the 
Perron-Frobenius operator) deterministic, while fuzzing up one of the characteristic functions in~(\ref{detTheta1}), then 
make the transfer operator act on the other one, that is still deterministic~\footnote{In~\ref{app:alter_evol}, we present a different but equivalent computation of Eq.~(\ref{noisyTheta1}).}, so as to write
%\begin{linenomath}
\begin{eqnarray}
\label{ThetaOm1}
\nonumber
\hspace{-10mm}
m\left(\Theta_{1,\Omega}(H)\right) &=& \int_M dy \, P_{H,\Omega}(y) \int_M dx\, \delta(y-f(x))\capchi_{M\setminus H}(x)  
 \\  \nonumber &=&  \int_M dy \, P_{H,\Omega}(y) \capchi_{M\setminus H}\left(f^{-1}(y)\right)  \\ \nonumber
&=& \int_M dy \, P_{H,\Omega}(y) \capchi_{M}\left(f^{-1}(y)\right) - \int_M dy \, P_{H,\Omega}(y) \capchi_{H}\left(f^{-1}(y)\right)
\\ \nonumber &=&  \int_M dy \, P_{H,\Omega}(y) -  \int_{f(H)} dy \, P_{H,\Omega}(y) \\ &=& \int_{M\setminus f(H)} P_{H,\Omega}(y)dy
\,,
\label{noisyTheta1}
\end{eqnarray}
%\end{linenomath}
assuming a unitary determinant for the Jacobian of the map $f$, as it is the case in the cat maps. 
At this point, $m(\cdot)$ denotes a probabilistic measure, and it is no longer an area.
In the previous evaluation, the probability of hopping into the hole only due to noise in one
iteration was given by
%\begin{linenomath}
\begin{equation}
P_{H,\Omega}(x) = \frac{m\left(H \bigcap W_\Omega(x)\right)}{m\left(W_\Omega(x)\right)}
\,,
\label{Pomega}
\end{equation}
%\end{linenomath}
where
%\begin{linenomath}
\begin{equation}
W_\Omega(x) = \left[ X -\frac{\Omega}{2}, X + \frac{\Omega}{2}\right)\,\times\,\left[ Y -\frac{\Omega}{2}, Y + \frac{\Omega}{2}\right)
\,.
\end{equation}
%\end{linenomath}
In order to proceed to the next step, formally
%\begin{linenomath}
\begin{equation}
m\left(\Theta_{2,\Omega}(H)\right) = \int_M dy \, \capchi_{H}(y) \int_M dx\, \delta_\Omega(y-f^2(x))\capchi_{M\setminus \left(H \bigcup f(H)\right)}(x)  
\,,
\label{mTheta2}
\end{equation}
%\end{linenomath}
assume that we may still separate noise and deterministic dynamics, so that, for instance, we apply the 
Perron-Frobenius operator twice to the characteristic function on the right, $\capchi_{M\setminus \left(H\bigcup f(H)\right)}(x)$,
and iterate the purely random dynamics twice on the left, that is: $i)$ transform $\capchi_{H}(y)\rightarrow P_{H,\Omega}(y)$, that is the probability to hop into the hole due to noise alone,   
%The first iteration yields the distribution $P_{H,\Omega}(x)$ 
as in Eq.~(\ref{Pomega}); $ii)$ apply noise again to $P_{H,\Omega}(y)$ , obtaining 
 %\begin{linenomath}
\begin{equation}
P_{H,\Omega}^{(1)}(y) = \frac{m\left(P_{H,\Omega}(y) \bigcap W_\Omega(y)\right)}{m\left(W_\Omega(y)\right)}
\,,
\label{PomegaP}
\end{equation}
%\end{linenomath}    
that is the probability to randomly hop into the hole in two iterations. Then Eq.~(\ref{mTheta2}) is,
more concisely,
%\begin{linenomath}
\begin{eqnarray}
\nonumber
m\left(\Theta_{2,\Omega}(H)\right) &=& \int_M dy P_{H,\Omega}^{(1)}(y)\capchi_{M\setminus \left(H\bigcup f(H)\right)}\left(f^{-2}(y)\right)     
\\ \nonumber &=&  \int_M dy P_{H,\Omega}^{(1)}(y) -  \int_{f^2(H)} dy P_{H,\Omega}^{(1)}(y) - \int_{f^3(H)} dy P_{H,\Omega}^{(1)}(y) \\ &=&
\int_{M\setminus[f^2(H)\bigcup f^3(H)]} P_{H,\Omega}^{(1)}(y)dy
\,.
\label{mTheta2bis}
\end{eqnarray}
%\end{linenomath}  
%I am not evaluating Eq.~(\ref{mTheta2}) explicitly here, yet
%%it gives the impression 
The previous expression for $m\left(\Theta_{2,\Omega}(H)\right)$ tells us that, even in this local model accounting for orthogonal stretching and
contraction (no folding), and noise, the exact computation of the escape rate is impractical. 
%and    
%the deterministic result 
%\begin{linenomath}
%\begin{equation}
%$m\left(\Theta_n(H)\right) = m\left(\Theta_1(H)\right)
%\,, 
%\hspace{0.2cm} 
%n>1$
%\end{equation}
%\end{linenomath}  
%no longer holds, if not approximately, in the presence of noise. %Now, because noise has an overall spreading effect on the dynamics,
%we may expect the integrals over the (positive definite) hopping probability densities $P_{H,\Omega}^{(n)}$ to systematically exceed their deterministic counterparts,
%that have characteristic functions as integrands. As a consequence (compare Eqs.~(\ref{ThetaOm1}) and~(\ref{mTheta2bis}) for example),
In order to work out an estimate for the escape rate, we make the approximation $m\left(\Theta_{n,\Omega}(H)\right)\approx m\left(\Theta_{1,\Omega}(H)\right)$ 
(that was an exact identity in the noiseless regime), and thus
%\begin{linenomath}
\begin{equation}
m\left(\Xi_{n,\Omega}(H)\right) \approx m(W_\Omega(x)) +  nm\left(\Theta_{1,\Omega}(H)\right)
\,,
\end{equation}
%\end{linenomath}  
and thus, for a sufficiently small opening [of size $l^2$, \textit{cf.} Eq.~(\ref{eq:Taylor series of gamma})],   
the escape rate for the map $f$ with noise is estimated as %and bounded by
\begin{eqnarray}
\label{eq:aprx of noisy escape rate}
\nonumber
\gamma(H,\Omega)&=&-\lim_{n\rightarrow\infty}\frac{1}{n}\ln m(M\setminus\Xi_{n, \Omega}(H))\\
&=&-\lim_{n\rightarrow\infty}\frac{1}{n}\ln (1-m(\Xi_{n, \Omega}(H))) \\ \nonumber 
&\simeq& \lim_{n\rightarrow\infty}\frac{m(\Xi_{n, \Omega}(H))}{n} \approx m\left(\Theta_{1,\Omega}(H)\right) 
\,.
%&=Q. 
\end{eqnarray}
%\begin{linenomath}
%\begin{equation}
%\frac{\gamma(H,\Omega)}{m(H)} \leq m\left(\Theta_{1,\Omega}(H)\right)
%\,.
%\end{equation}
In what follows, we compute the above estimate for $\gamma(H,\Omega)$ in the local model perturbed by noise of
varying amplitude, and then compare it with the numerically obtained escape rate of the cat map. 
%\end{linenomath}  

%\subsection{Escape rate for the local model}
%\label{sec:escape_rate_local_model}

%Letting $P(X, Y)$ be the probability density with which the orbits at $(X,Y)$ enter the hole only due to noise, we can express the area occupied by the set $\Theta_{n, \Omega}(H)$ as 
%\begin{equation}
%\label{eq:def_q}
%m(\Theta_{n, \Omega}(H))=\begin{array}{cc}
%\vspace{2mm}
%\displaystyle 
%m(H), & n=0,\\ %\vspace{2mm}
%\displaystyle 
%\int_{M\setminus f(H)} P(X, Y)dXdY, %=:Q.  
%& n>0 .
%\end{array}
%\end{equation}
First, we need to evaluate  the probability density of hopping into the hole due to noise, $P_{H,\Omega}(x)$.
In this specific model, the local stable and unstable directions are split into vertical and horizontal directions, and noises $\omega_{n}^{(X)}$ and $\omega_{n}^{(Y)}$ are applied independently. Hence,  we may decompose $P_{H,\Omega}(x)= P(X, Y) = P(X)P(Y)$, where $P(X)$ and $P(Y)$ are identical one-dimensional probability densities. 

Because of the factorization, we may then just find $P(X)$ by addressing the one-dimensional problem: 
suppose the points are uniformly distributed on the line with the hole $H_X:=(-l/2, l/2)$. 
Denote the points that may enter the hole $H_X$ by applying uniform random noise with strength $\Omega$ by 
\begin{equation}
\label{eq:W}
W_{\Omega}(X):=\left[X-\frac{\Omega}{2}, X+\frac{\Omega}{2}\right). 
\end{equation}
Then the probability density with which the orbits enter the hole $H_X$ is expressed as
\begin{equation}
\label{eq:p}
P(X)=\frac{m(H_X\cap W_{\Omega}(X))}{m(W_{\Omega}(X))}=\frac{m(H_X\cap W_{\Omega}(X))}{\Omega}
\,.
\end{equation}
%where $m(\cdot)$ denotes the length of the interval. 
Depending on the noise strength $\Omega$ and the position $X$, the calculation yields different outcomes. 
If $\Omega<l$, we have  
\begin{enumerate}
\item $X+\Omega/2<-l/2$: there is no chance for the points to enter the hole, so 
\begin{equation}
m(H_X\cap W_{\Omega}(X))=0
\,. %trivially holds. 
\end{equation}
\item %If the conditions 
$X-\Omega/2<-l/2$ and $-l/2\leq X+\Omega/2$: 
%are satisfied, 
we find 
\begin{eqnarray}
\label{eq:mHW2}
m(H_X\cap W_{\Omega}(X))&=&X+\frac{\Omega}{2}-\left(-\frac{l}{2}\right)\nonumber \\
&=&X+\frac{l+\Omega}{2}. 
%,\qquad -\frac{l+\Omega}{2}\leq X<-\frac{l-\Omega}{2}
\end{eqnarray}
\item  $-l/2\leq X-\Omega/2$ and $X+\Omega/2<l/2$: 
then $W_{\Omega}(X)$ is a subset of the hole, which leads to 
\begin{equation}
\label{eq:mHW3}
m(H_X\cap W_{\Omega}(X))=\Omega. 
%\qquad -\frac{l-\Omega}{2}\leq X<\frac{l-\Omega}{2}
\end{equation}
\item $X-\Omega/2<l/2$ and $l/2\leq X+\Omega/2$: 
we have 
\begin{eqnarray}
\label{eq:mHW4}
m(H_X\cap W_{\Omega}(X))&=&\frac{l}{2}-\left(X-\frac{\Omega}{2}\right) \nonumber \\
&=&-X+\frac{l+\Omega}{2}. 
%,\qquad \frac{l-\Omega}{2}\leq X<\frac{l+\Omega}{2}
\end{eqnarray}
\item $l/2\leq X-\Omega/2$: the hole $H_X$ and 
$W_{\Omega}(X)$ do not have any intersection, so we have 
\begin{equation}
\label{eq:mHW5}
m(H_X\cap W_{\Omega}(X))=0. 
\end{equation}
%There arguments lead to the formula (\ref{eq:1-dimensional1}). 
%We can similarly derive the formula (\ref{eq:1-dimensional2}) for $l \le \Omega$. 
%As shown in Appendix \ref{app:derivation_one_dimensional}, for $\Omega < l$, 
%the probability density $P(X)$, with which the points uniformly distributed on the line enter the hole $H=(-l/2, l/2)$ under the uniform random noise $\omega_{n}\in[-\Omega/2, \Omega/2)$, is obtained as
\end{enumerate}
We may summarize the above results as follows 
\begin{equation}
\label{eq:1-dimensional1} 
P(X)=\left\{ \begin{array}{ll}
\vspace{2mm}
\displaystyle 
0, &  \hspace{5mm} \displaystyle X< - (l+ \Omega)/2, \\ %-\frac{l+\Omega}{2},\\
\vspace{2mm}
\displaystyle 
\frac{1}{\Omega}X+\frac{l+\Omega}{2\Omega}, & 
\hspace{5mm} \displaystyle 
- (l+\Omega)/2 \leq X < - (l-\Omega)/2, \\
%\frac{l+\Omega}{2}\leq X<-\frac{l-\Omega}{2},\\
\vspace{2mm}
\displaystyle 
1, & \hspace{5mm} 
-(l-\Omega)/2 \leq X< (l-\Omega)/2, \\
%-\frac{l-\Omega}{2}\leq X<\frac{l-\Omega}{2},\\
\vspace{2mm}
\displaystyle 
-\frac{1}{\Omega}X+\frac{l+\Omega}{2\Omega}, & 
\hspace{5mm} \displaystyle 
(l-\Omega)/2 \leq X< (l+\Omega)/2, \\
%\frac{l-\Omega}{2}\leq X<\frac{l+\Omega}{2},\\
\vspace{2mm}
\displaystyle 
0, & \hspace{5mm} \displaystyle 
(l+\Omega)/2 \leq X.
%\frac{l+\Omega}{2}\leq X.
\end{array}
\right.
\end{equation}
Similarly, for the $\Omega > l$ case, an analogous calculation yields
\begin{equation}
\label{eq:1-dimensional2}
P(X)= \left\{ \begin{array}{ll}
\vspace{2mm}
\displaystyle 
0, & \hspace{5mm} \displaystyle  X<- (l+\Omega)/2, \\
%\\X<-\frac{l+\Omega}{2},\\
\vspace{2mm}
\displaystyle 
\frac{1}{\Omega}X+\frac{l+\Omega}{2\Omega}, & 
\hspace{5mm} \displaystyle -(l+\Omega)/2 \leq X<-(\Omega-l)/2, \\
\vspace{2mm}
\displaystyle %-\frac{l+\Omega}{2}\leq X<-\frac{\Omega-l}{2},\\
\frac{l}{\Omega}, & 
\hspace{5mm} \displaystyle
-(\Omega-l)/2 \leq X< (\Omega-l)/2,\\
%-\frac{\Omega-l}{2}\leq X<\frac{\Omega-l}{2},\\
\vspace{2mm}
\displaystyle 
-\frac{1}{\Omega}X+\frac{l+\Omega}{2\Omega}, & 
\hspace{5mm} \displaystyle
(\Omega-l)/2 \leq X< (l+\Omega)/2,\\
%\frac{\Omega-l}{2}\leq X<\frac{l+\Omega}{2},\\
\vspace{2mm}
\displaystyle 
0, & \hspace{5mm} \displaystyle
(l+\Omega)/2 \leq X.
%\frac{l+\Omega}{2}\leq X.
\end{array}
\right.
\end{equation}

%%%%%%%%%%%%%%%%%%%%%%%%%%%%%%%%%%%%%%%%%%
%%%%%%%%%%%%%%%%%       Fig. 9         %%%%%%%%%%%%%%%%
%%%%%%%%%%%%%%%%%%%%%%%%%%%%%%%%%%%%%%%%%%
\begin{figure}
\begin{minipage}{1.0\hsize}
\begin{center}
\vspace{10mm}
\includegraphics[width = 6.0cm,bb= 0 0 461 346]{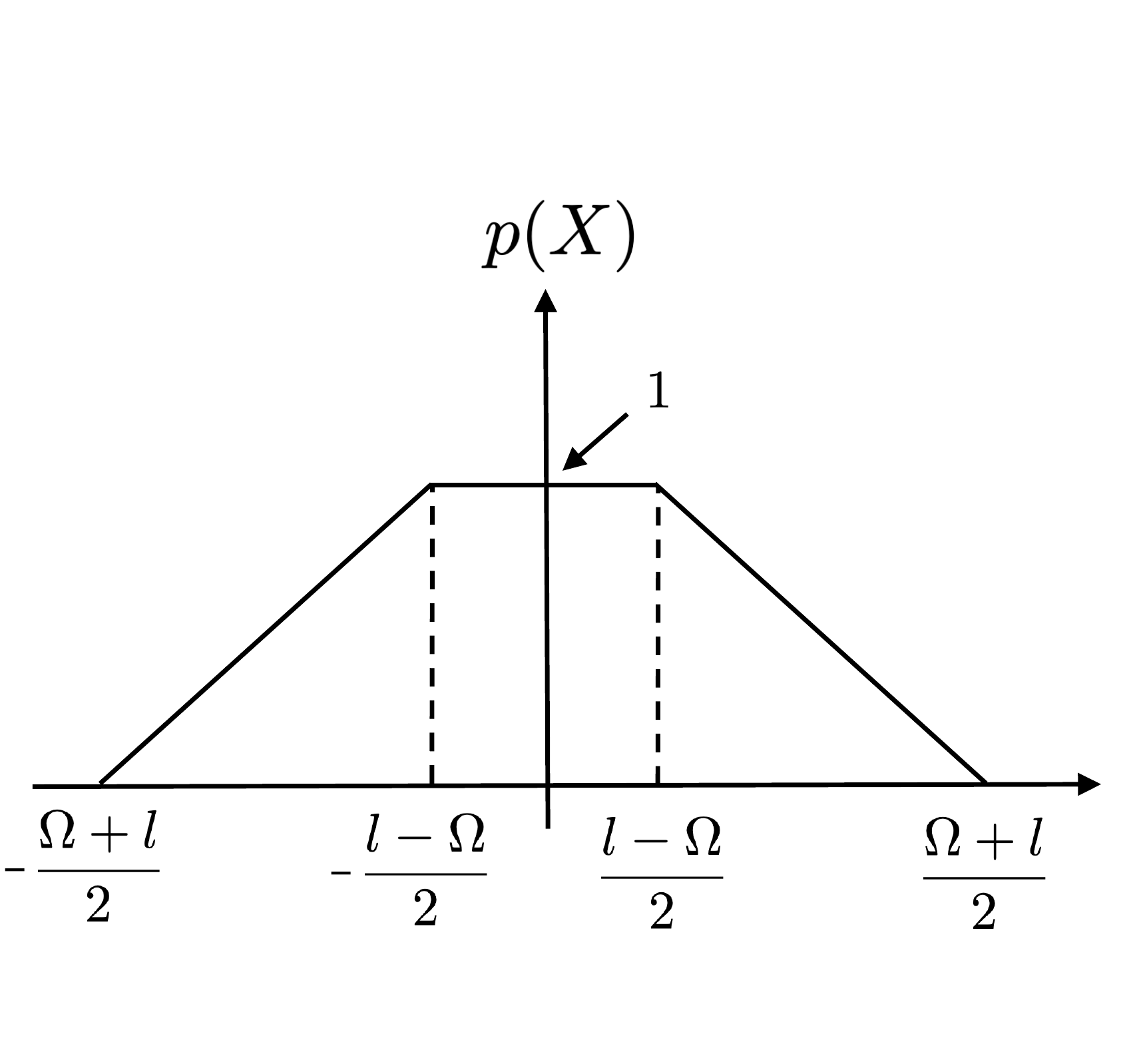}
\hspace{5mm}
\includegraphics[width = 6.0cm,bb= 0 0 461 346]{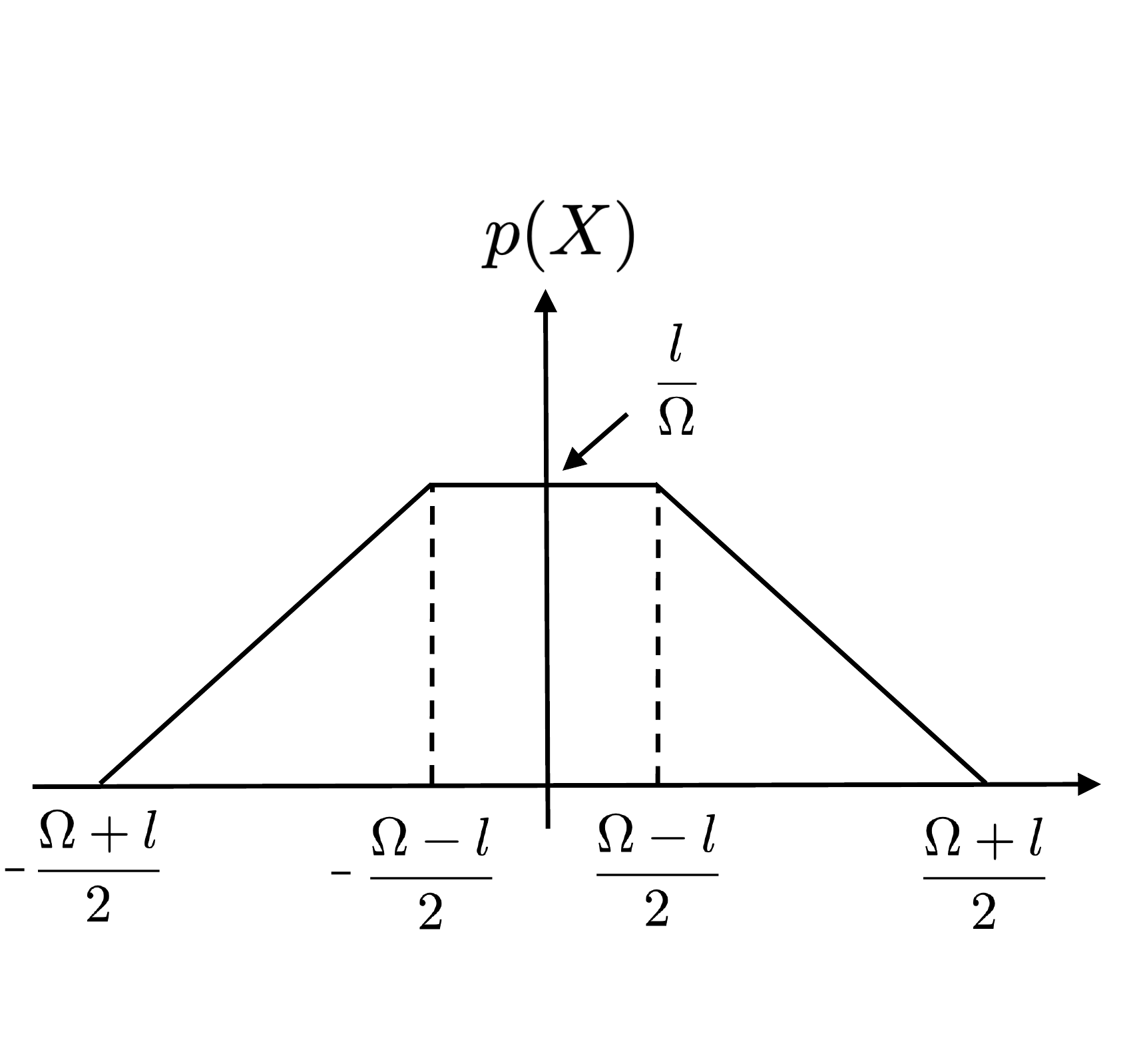}
\end{center}
\end{minipage}
\caption{
Probability density $P(X)$ as defined by Eq.~(\ref{eq:p}) in the cases of (a) $\Omega < l$ and (b) $l < \Omega$. 
$P(Y)$ has the same form as $P(X)$.
}
\label{fig:fig7}
\end{figure} 
%%%%%%%%%%%%%%%%%%%%%%%%%%%%%%%%%%%%%%%%%
%%%%%%%%%%%%%%%%%%%%%%%%%%%%%%%%%%%%%%%%%
%%%%%%%%%%%%%%%%%%%%%%%%%%%%%%%%%%%%%%%%%

%In the derivation of (\ref{eq:Taylor series of gamma}), for sufficiently small $l$, 
%the escape rate for the map $F_n$ with noise is given as
%\begin{eqnarray}
%\label{eq:aprx of noisy escape rate}
%\gamma(H,\Omega)&=-\lim_{n\rightarrow\infty}\frac{1}{n}\ln m(M\setminus\Xi_{n, \Omega}(H))\\
%&=-\lim_{n\rightarrow\infty}\frac{1}{n}\ln (1-m(\Xi_{n, \Omega}(H)))
%&\simeq \lim_{n\rightarrow\infty}\frac{m(\Xi_{n, \Omega}(H))}{n} . 
%\\
%&=Q. 
%\end{eqnarray}
%Hence, our task is reduced to evaluating the right-hand side defined in (\ref{eq:def_q}).

We may evaluate the measure of the set of trajectories that escape in one iteration, $m(\Theta_{1,\Omega}(H))$, from Eq.~(\ref{ThetaOm1}).
For this purpose, we have to consider the relation between the interval $[-\lambda l/2, \lambda l/2]$ and the interval specifying the condition (\ref{eq:1-dimensional1}) (or (\ref{eq:1-dimensional2})) in the $X$-direction. Likewise, the relation between the interval $[-l/2\lambda, l/2\lambda]$ and the interval specifying the condition (\ref{eq:1-dimensional1}) (or (\ref{eq:1-dimensional2})) in the $Y$-direction should be considered. 

In order to evaluate the integral~(\ref{ThetaOm1}), we need to determine the relative positions of the support of the probability density $P(X)$, the boundaries of the opening $H_X$ and of its image $f(H_X)$. 
Since $\lambda>1$ is assumed, the inequalities shown below hold depending 
on the value of $\lambda$. \newline For $1 < \lambda \le 1 + \sqrt{2}$,  
\begin{equation}
0 < 1 - \frac{1}{\lambda} < 1 <  \lambda -1  \le 1 + \frac{1}{\lambda} < \lambda + 1. 
\end{equation}
For $1+ \sqrt{2} < \lambda$, 
\begin{equation}
0< 1 - \frac{1}{\lambda} < 1 < 1 + \frac{1}{\lambda} < \lambda -1 < \lambda + 1, 
\end{equation}
holds.
Using these relations, we identify six different ranges of noise strength, that determine as many limits of integration in~(\ref{ThetaOm1}), for each of the following two cases:
%\bn

\bigskip
\noindent
For $1 < \lambda \le 1 + \sqrt{2}$, 
\vspace{-2mm}
\begin{tabbing}
%\n
\={\rm (i)} ~~~~~~\= $0<\Omega<l(1-1/\lambda)$.  \\
\>{\rm (ii)}  \> $l(1-1/\lambda) < \Omega<l$. \\
\>{\rm (iii)}  \> $l\leq\Omega<l(\lambda-1)$. \\
\>{\rm (iv)}  \> $l(\lambda-1) < \Omega<l(1+1/\lambda)$. \\
\>{\rm (v)}  \> $l(1+1/\lambda) < \Omega<l(\lambda+1)$. \\
\>{\rm (vi)}  \> $l(\lambda+1)  < \Omega$. 
\end{tabbing}

\bigskip
\noindent
For $1+ \sqrt{2} < \lambda$,  
\vspace{-2mm}
\begin{tabbing}
%\n
\={\rm (i)} ~~~~~~\= $0<\Omega<l(1-1/\lambda)$.  \\
\>{\rm (ii)}  \> $l(1-1/\lambda) < \Omega<l$. \\
\>{\rm (iii)}  \> $l\leq\Omega<l(1+1/\lambda)$. \\
\>{\rm (iv)}  \> $l(1+1/\lambda) < \Omega<l(\lambda-1)$. \\
\>{\rm (v)}  \> $l(\lambda-1) < \Omega<l(\lambda+1)$. \\
\>{\rm (vi)}  \> $l(\lambda+1)< \Omega$. 
\end{tabbing}

\begin{figure}
\begin{minipage}{1.0\hsize}
\begin{center}
\vspace{20mm}
\includegraphics[width = 5.0cm,bb= 0 0 461 346]{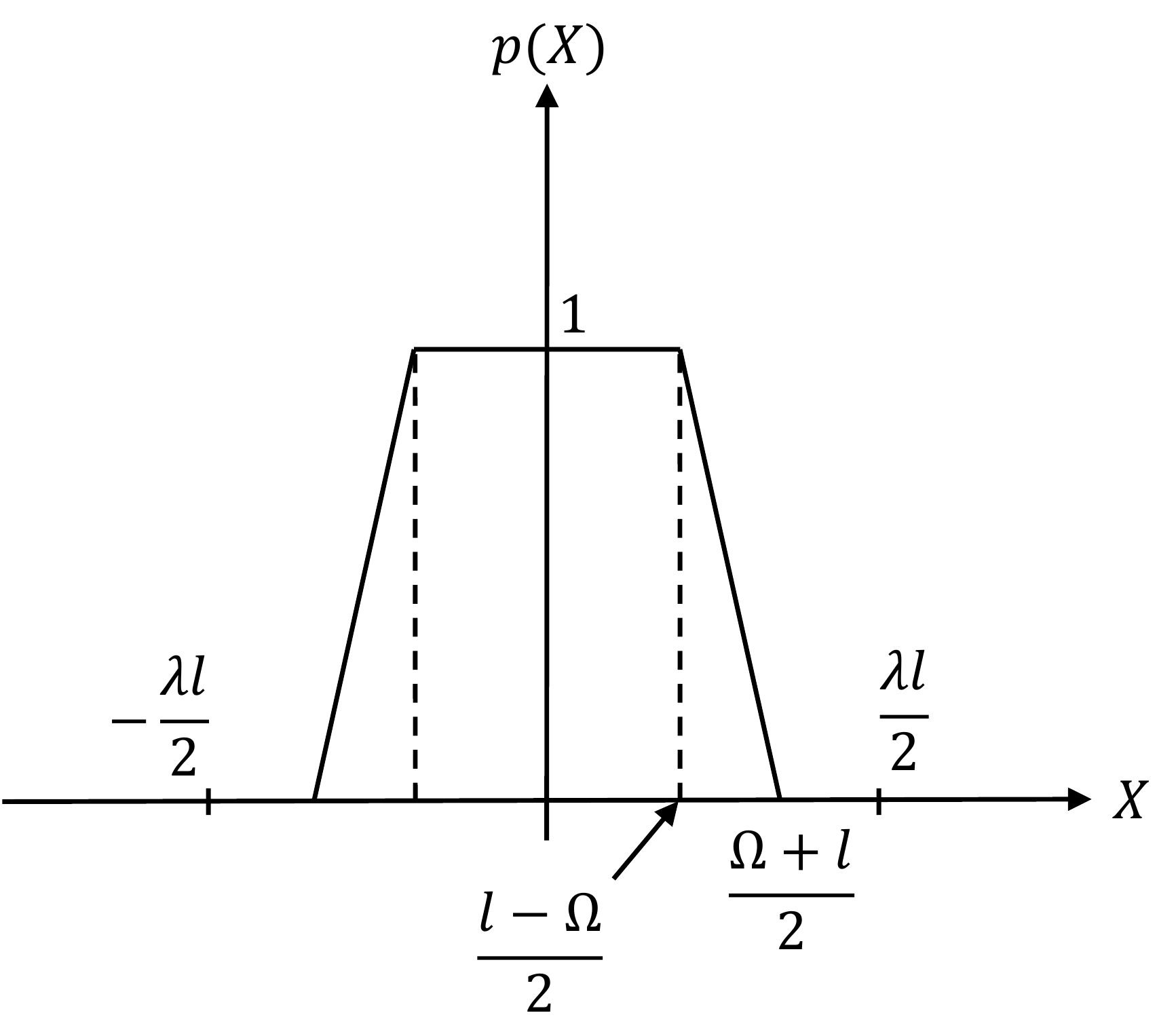}
\hspace{14mm}
\includegraphics[width = 5.0cm,bb= 0 0 461 346]{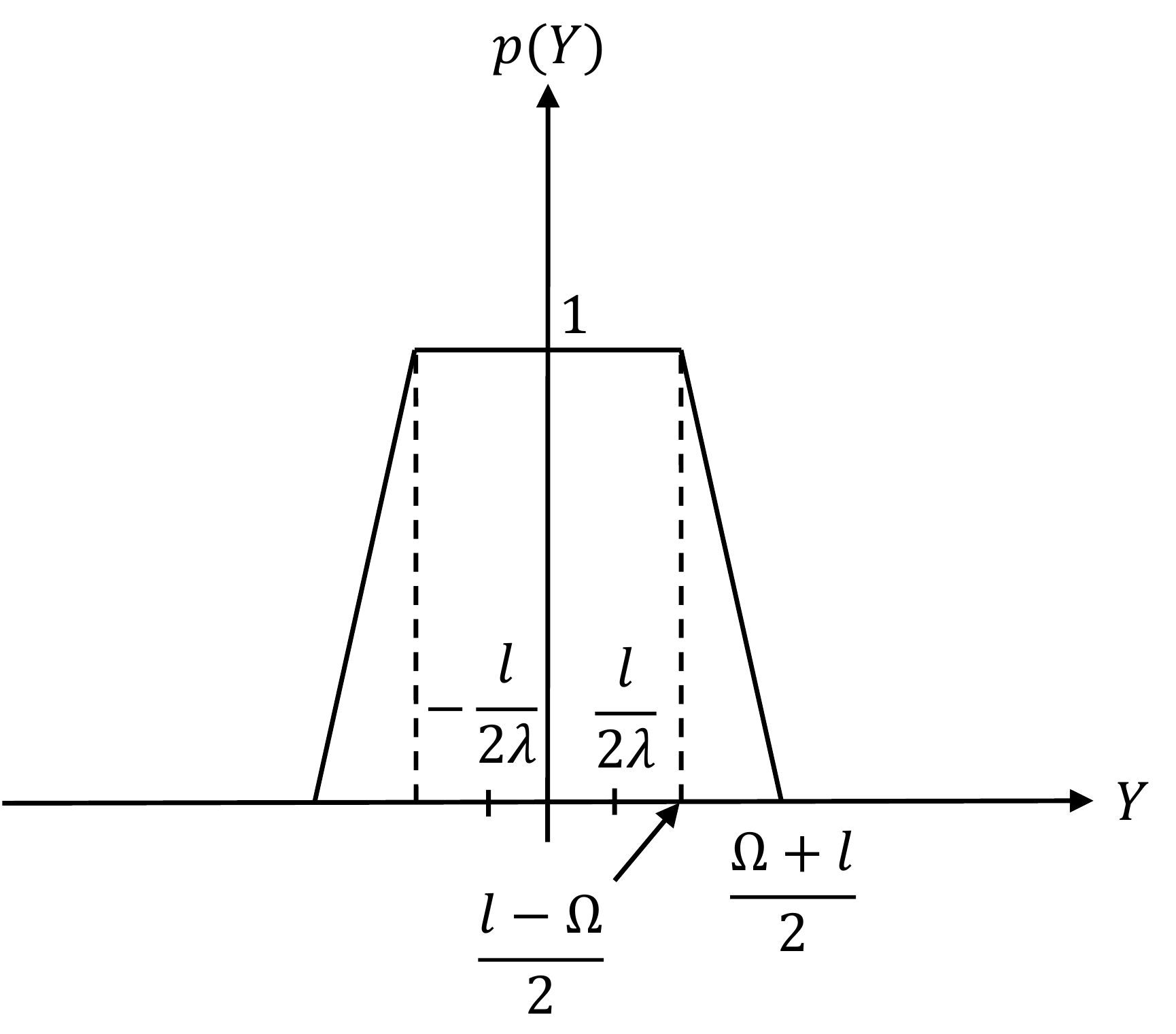}
\end{center}
\end{minipage}
\caption{
Probability densities (a) $P(X)$ and (b) $P(Y)$ as defined by Eq.~(\ref{eq:p}), and position of the boundaries of $f(H_X)$ in the case $0<\Omega < l(1-1/\lambda)$. 
}
\label{fig:proba}
\end{figure} 
%%%%%%%%%%%%%%%%%%%%%%%%%%%%%%%%%%%%%%%%%
%%%%%%%%%%%%%%%%%%%%%%%%%%%%%%%%%%%%%%%%%
%%%%%%%%%%%%%%%%%%%%%%%%%%%%%%%%%%%%%%%%%

In each case, we can evaluate the approximation $m(\Theta_{1,\Omega}(H))$ to the escape rate $\gamma(H,\Omega)$ as follows:

%\bn
For $1 < \lambda \le 1 + \sqrt{2}$,  %from Eq.~(\ref{ThetaOm1}).  
and in the noise amplitude range $0<\Omega<l(1-1/\lambda)$, %$\gamma(H,\Omega)$ is estimated as
the integral of the probability densities over the torus $M$ is performed following Eq.~(\ref{eq:1-dimensional1}) (see Fig.~\ref{fig:fig7}), whereas
the integral over $f(H)$ must take into account the relative positions of the boundaries of the mapped hole with the different intervals of the
piecewise linear $P(X)$, $P(Y)$ (see Fig.~\ref{fig:proba}). The outcome is 
%\bn
\begin{eqnarray}
\label{eq:Q1}
\nonumber
\hspace{-10mm}
\gamma(H,\Omega) &\approx& m(\Theta_{1,\Omega}(H)) = \int\int_{M\setminus f(H)} P(X)P(Y)dXdY \\ \nonumber
&=&\int\int_{M} P(X)P(Y)dXdY-\int\int_{f(H)} P(X)P(Y)dXdY \\ \nonumber
&=&\int_{-\frac{l+\Omega}{2}}^{\frac{l+\Omega}{2}}P(X)dX\int_{-\frac{l+\Omega}{2}}^{\frac{l+\Omega}{2}}P(Y)dY - \int_{-\frac{\lambda l}{2}}^{\frac{\lambda l}{2}}P(X)dX\int_{-\frac{l}{2\lambda}}^{\frac{l}{2\lambda}}P(Y)dY \\
%&=l^{2}-l\cdot\frac{l}{\lambda}\notag\\
&=&l^{2}\left(1-\frac{1}{\lambda}\right). 
\end{eqnarray}
Since $m(H)=m(H_X\times H_Y) = l^{2}$, the escape rate $\gamma(H)$ takes the same form 
as the formula (\ref{eq:escape_rate}), 
meaning that noise does not affects the escape rate 
in such a small noise strength region.  In a similar manner, we calculate 
the escape rate for other cases (ii) - (vi), %(details in~\ref{app:restofcalc}), 
and the results are summarized as 
%\begin{widetext}
\begin{equation}
\label{eq:gamma1}
\hspace{-25mm}
\gamma(H,\Omega)\approx\left\{
\begin{array}{ll}
\vspace{2mm}
\displaystyle 
l^{2}\left(1-1/\lambda \right), 
& 0<\Omega<l\left(1-1/\lambda \right), \\ 
\vspace{2mm}
\displaystyle 
\frac{l}{4\Omega}\left(l-\frac{l}{\lambda}+\Omega\right)^{2} , 
& l\left(1-1/\lambda \right)\leq\Omega<l(\lambda-1) \,, \\ %l(\lambda-1), \\
\vspace{2mm}
\displaystyle 
\frac{l}{4\Omega}\left( \left(l-l/\lambda+\Omega\right)^{2}+\left(l-\lambda l+\Omega\right)^{2}\right)
&   \, %l\leq \Omega < l(\lambda -1) 
\\ \vspace{2mm}
\displaystyle 
-\frac{1}{16\Omega^{2}}\left(l-l/\lambda+\Omega\right)^{2}\left(l-\lambda l+\Omega\right)^{2} , 
& l(\lambda-1)\leq\Omega<l\left(1+1/\lambda \right), \\
\vspace{2mm}
\displaystyle 
l^{2}\left(1-\left(\frac{l}{\lambda\Omega}-\frac{1}{4\lambda\Omega^{2}}\left(l-\lambda l+\Omega\right)^{2}\right)\right), 
& l\left(1+1/\lambda \right)\leq\Omega<l(\lambda+1), \\
\vspace{2mm}
\displaystyle 
l^{2}\left(1-\frac{l^{2}}{\Omega^{2}}\right),  & ~~~~~\Omega\geq l(\lambda+1) .
\end{array}
\right.
\end{equation}
%\end{widetext}

%\n
For $1+ \sqrt{2} < \lambda$, we can similarly obtain 
%\begin{widetext}
\begin{equation}
\label{eq:gamma2}
\hspace{-25mm}
\gamma(H,\Omega)\approx\left\{
\begin{array}{ll}
\vspace{2mm}
\displaystyle 
l^{2}\left(1-1/\lambda\right),  & 0<\Omega<l\left(1-1/\lambda\right)\,, \\
\vspace{2mm}
\displaystyle 
\frac{l}{4\Omega}\left(l-\frac{l}{\lambda}+\Omega\right)^{2},  & l\left(1-1/\lambda\right)\leq\Omega<l\left(1+1/\lambda\right)\,, \\
\vspace{2mm}
\displaystyle 
l^{2}\left(1-\frac{l}{\lambda\Omega}\right),  & l\left(1+1/\lambda\right)\leq\Omega<l(\lambda-1)\,, \\
\vspace{2mm}
\displaystyle 
l^{2}\left(1-\left(\frac{l}{\lambda\Omega}-\frac{1}{4\lambda\Omega^{2}}\left(l-\lambda l+\Omega\right)^{2}\right)\right)\,, & l(\lambda-1)\leq\Omega<l(\lambda+1)\,, \\
\vspace{2mm}
\displaystyle 
l^{2}\left(1-\frac{l^{2}}{\Omega^{2}}\right) , & ~~~~~\Omega\geq l(\lambda+1)\,.
\end{array}
\right.
\end{equation}
%\end{widetext}

%%%%%%%%%%%%%%%%%%%%%%%%%%%%%%%%%%%%%%%%%%
%%%%%%%%%%%%%%%%%       Fig. 11         %%%%%%%%%%%%%%%%
%%%%%%%%%%%%%%%%%%%%%%%%%%%%%%%%%%%%%%%%%%
\begin{figure}
\begin{minipage}{1.1\hsize}
\begin{center}
\includegraphics[width = 6.7cm,bb= 0 0 461 346]{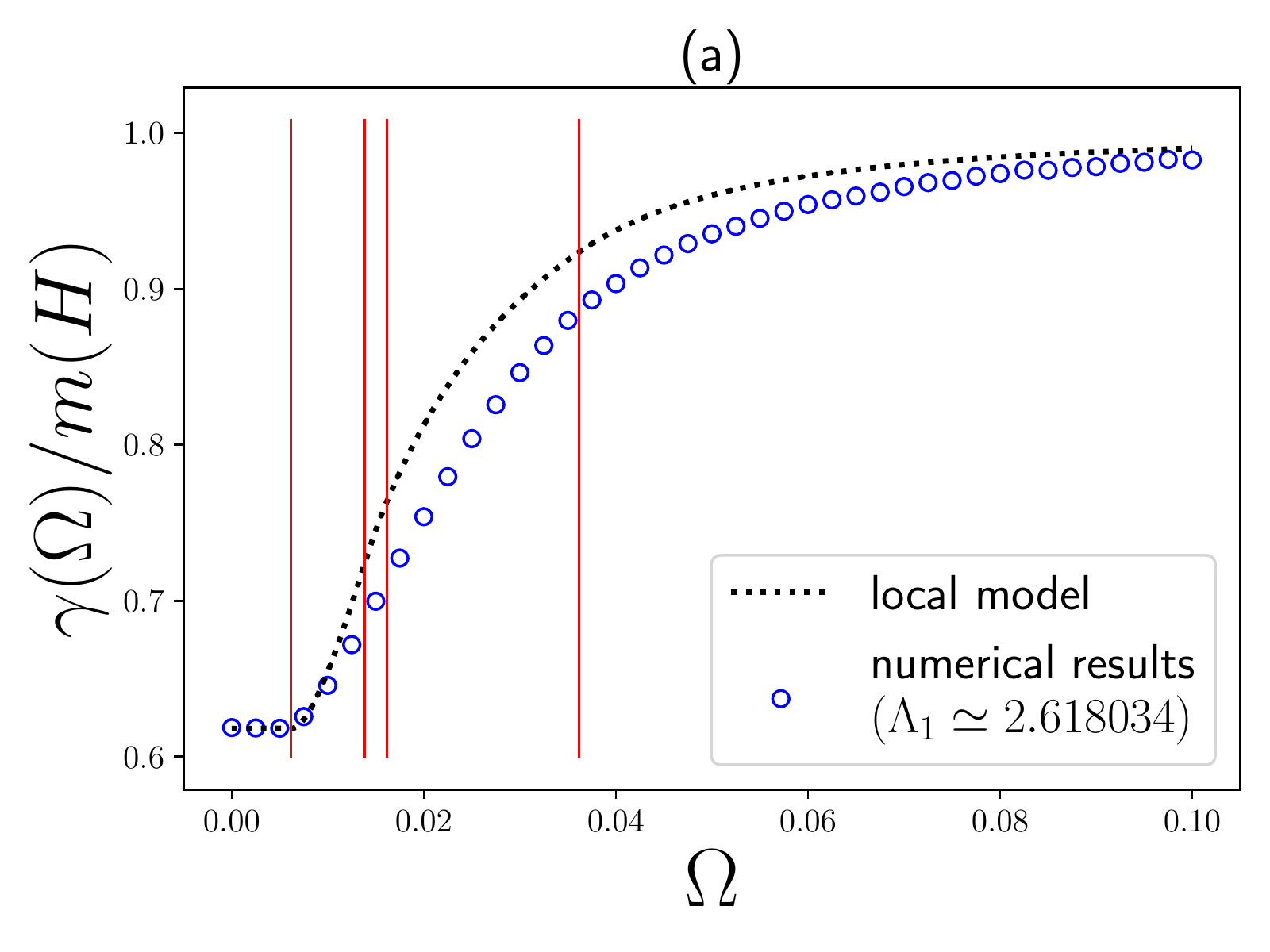}
\includegraphics[width = 6.7cm,bb= 0 0 461 346]{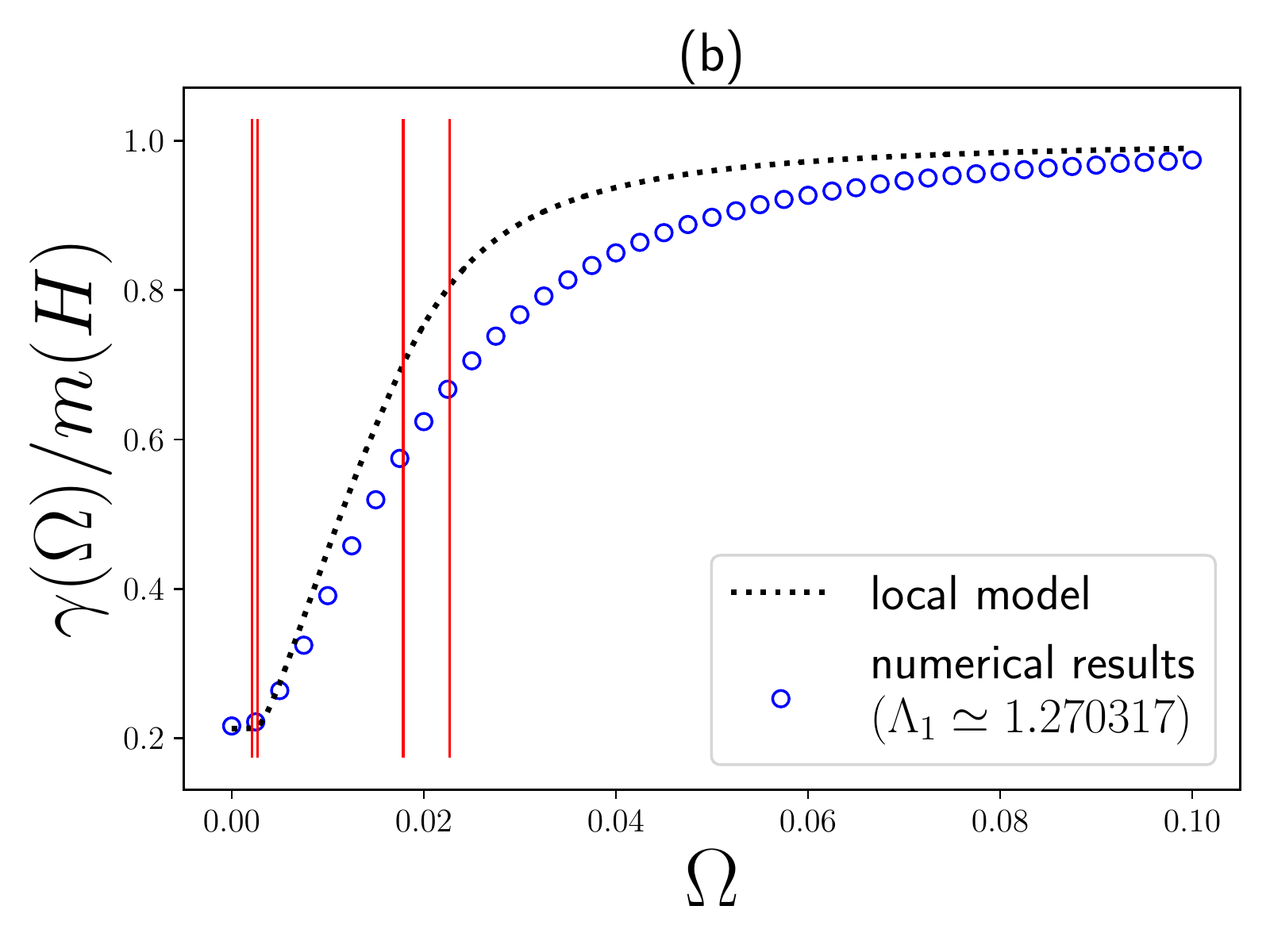}
\end{center}
\end{minipage}
\caption{
Comparison between the numerical results of the cat map and the theoretical prediction based on the local model. The escape rate is evaluated with the hole $H$ containing the fixed point at the center. Blue dots and black dotted line represent the results obtained by direct numerical calculations and the prediction, respectively. 
(a) Noisy cat map ($\epsilon=0$) with $1 < \lambda \le 1 + \sqrt{2}$, and (b) noisy perturbed cat map with $\epsilon=0.15$, $\nu=2$, with $1+ \sqrt{2} < \lambda$. 
The red vertical lines indicate the interval specified in Eqs.~(\ref{eq:gamma1}) and~(\ref{eq:gamma2}). 
}
\label{fig:fig8}
\end{figure}
%%%%%%%%%%%%%%%%%%%%%%%%%%%%%%%%%%%%%%%%%
%%%%%%%%%%%%%%%%%%%%%%%%%%%%%%%%%%%%%%%%%
%%%%%%%%%%%%%%%%%%%%%%%%%%%%%%%%%%%%%%%%%

%%%%%%%%%%%%%%%%%%%%%%%%%%%%%%%%%%%%%%%%%%
%%%%%%%%%%%%%%%%%       Fig. 12        %%%%%%%%%%%%%%%%
%%%%%%%%%%%%%%%%%%%%%%%%%%%%%%%%%%%%%%%%%%
\begin{figure}
\begin{minipage}{1.2\hsize}
\begin{center}
\includegraphics[width = 5.0cm,bb= 0 0 461 346]{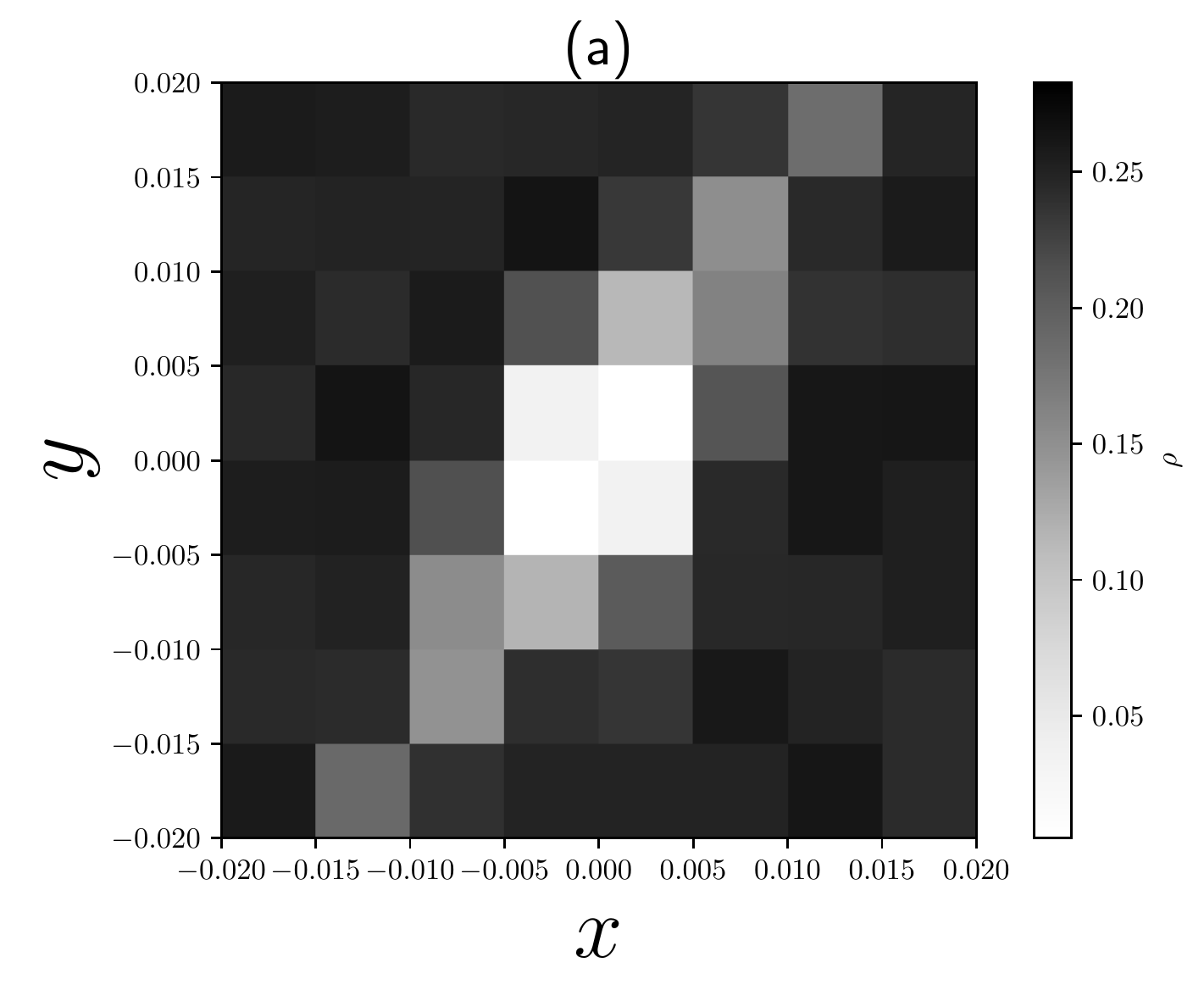}
\includegraphics[width = 5.0cm,bb= 0 0 461 346]{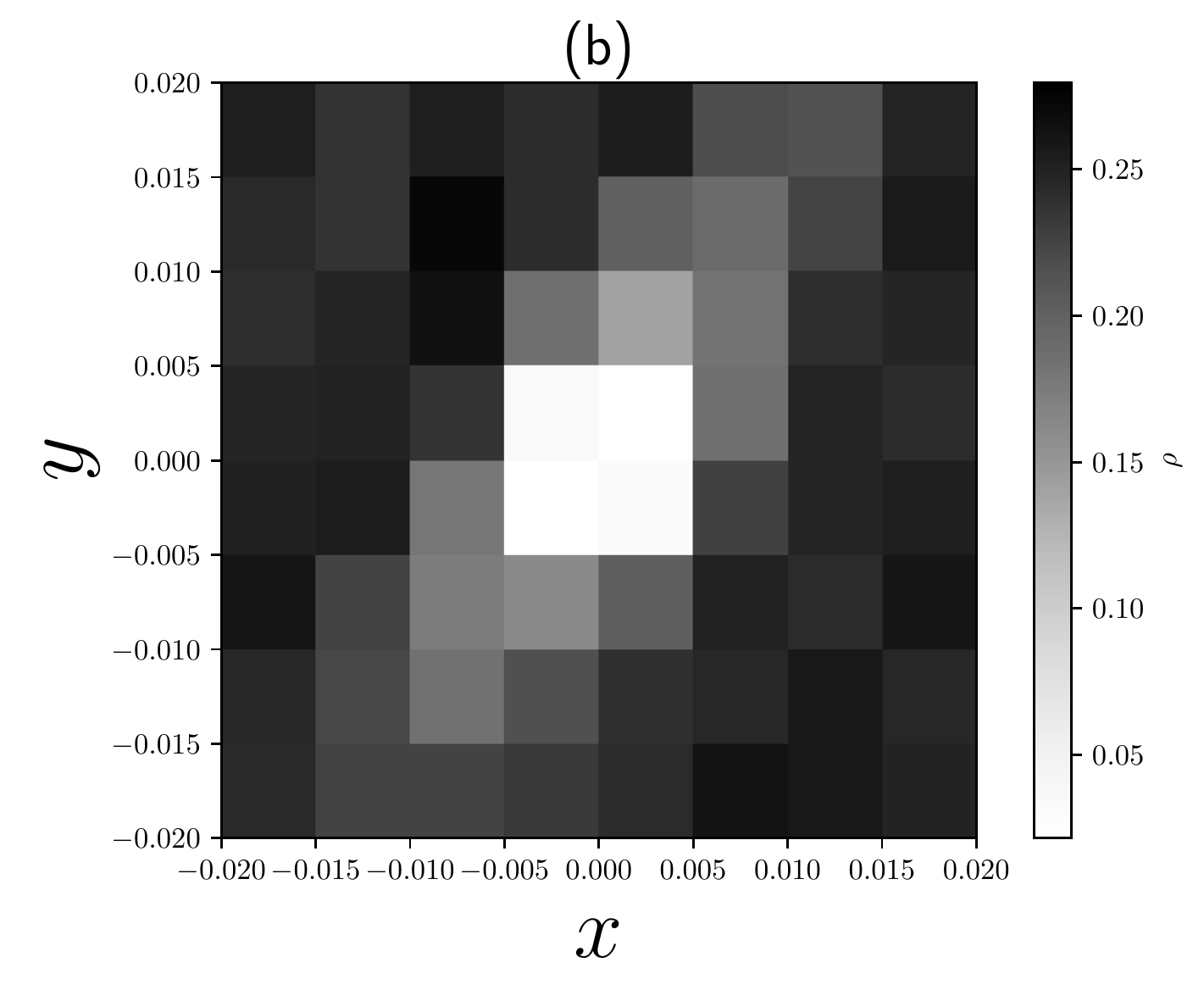}
\includegraphics[width = 5.0cm,bb= 0 0 461 346]{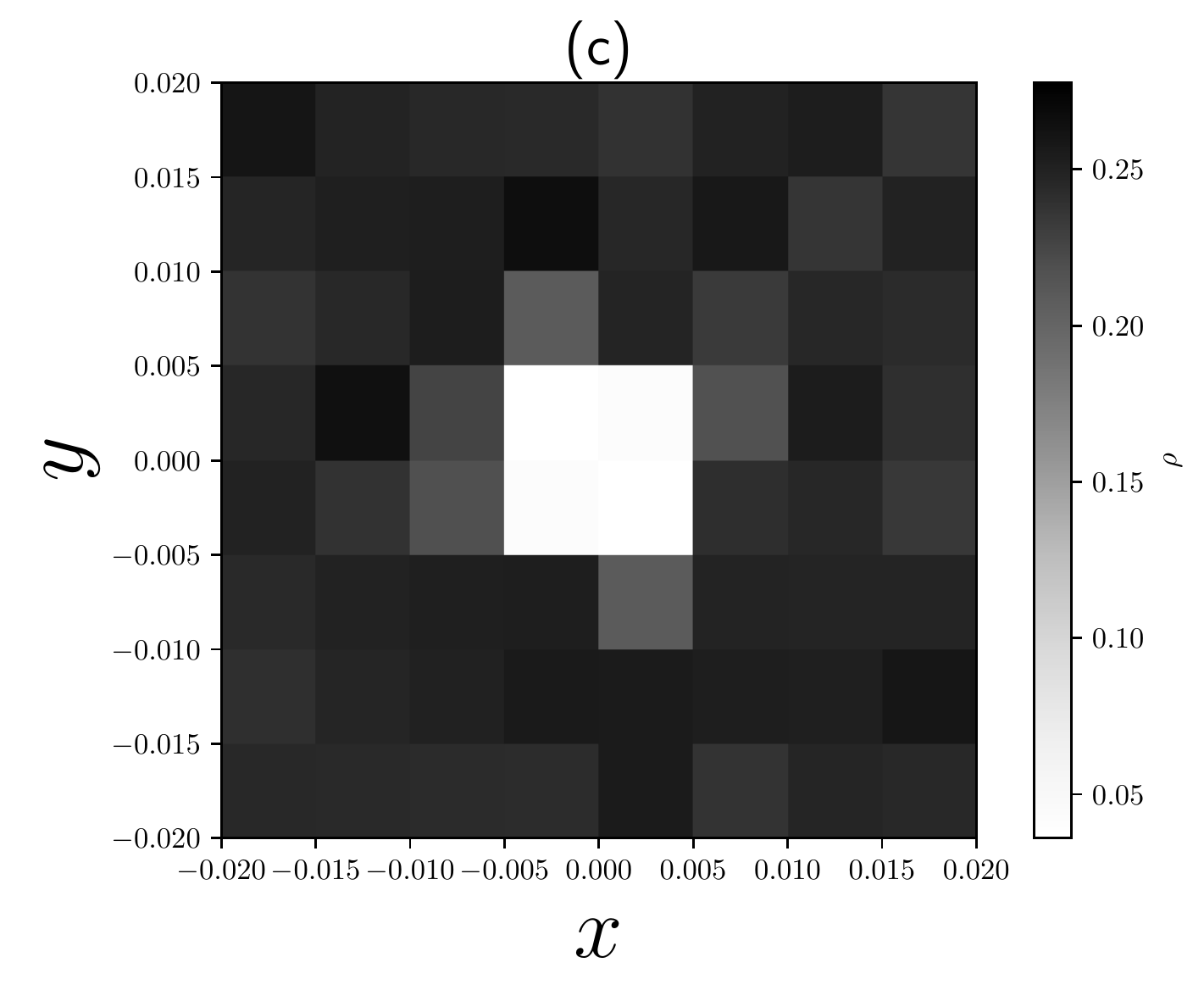}
\end{center}
\end{minipage}
\caption{
The density of orbits around the hole at $n=100$ in the case of the noisy cat map, from a uniform initial density.
The noise strength is respectively taken as (a) $\Omega=0$,  (b) $\Omega=0.01$, and (c) $\Omega=0.1$. 
The hole contains the fixed point $(0, 0)$.
}
\label{fig:fig9}
\end{figure}
%%%%%%%%%%%%%%%%%%%%%%%%%%%%%%%%%%%%%%%%%
%%%%%%%%%%%%%%%%%%%%%%%%%%%%%%%%%%%%%%%%%
%%%%%%%%%%%%%%%%%%%%%%%%%%%%%%%%%%%%%%%%%

%%%%%%%%%%%%%%%%%%%%%%%%%%%%%%%%%%%%%%%%%%
%%%%%%%%%%%%%%%%%       Fig. 13        %%%%%%%%%%%%%%%%
%%%%%%%%%%%%%%%%%%%%%%%%%%%%%%%%%%%%%%%%%%
\begin{figure}
\begin{minipage}{1.1\hsize}
\begin{center}
\includegraphics[width = 6.6cm,bb= 0 0 461 346]{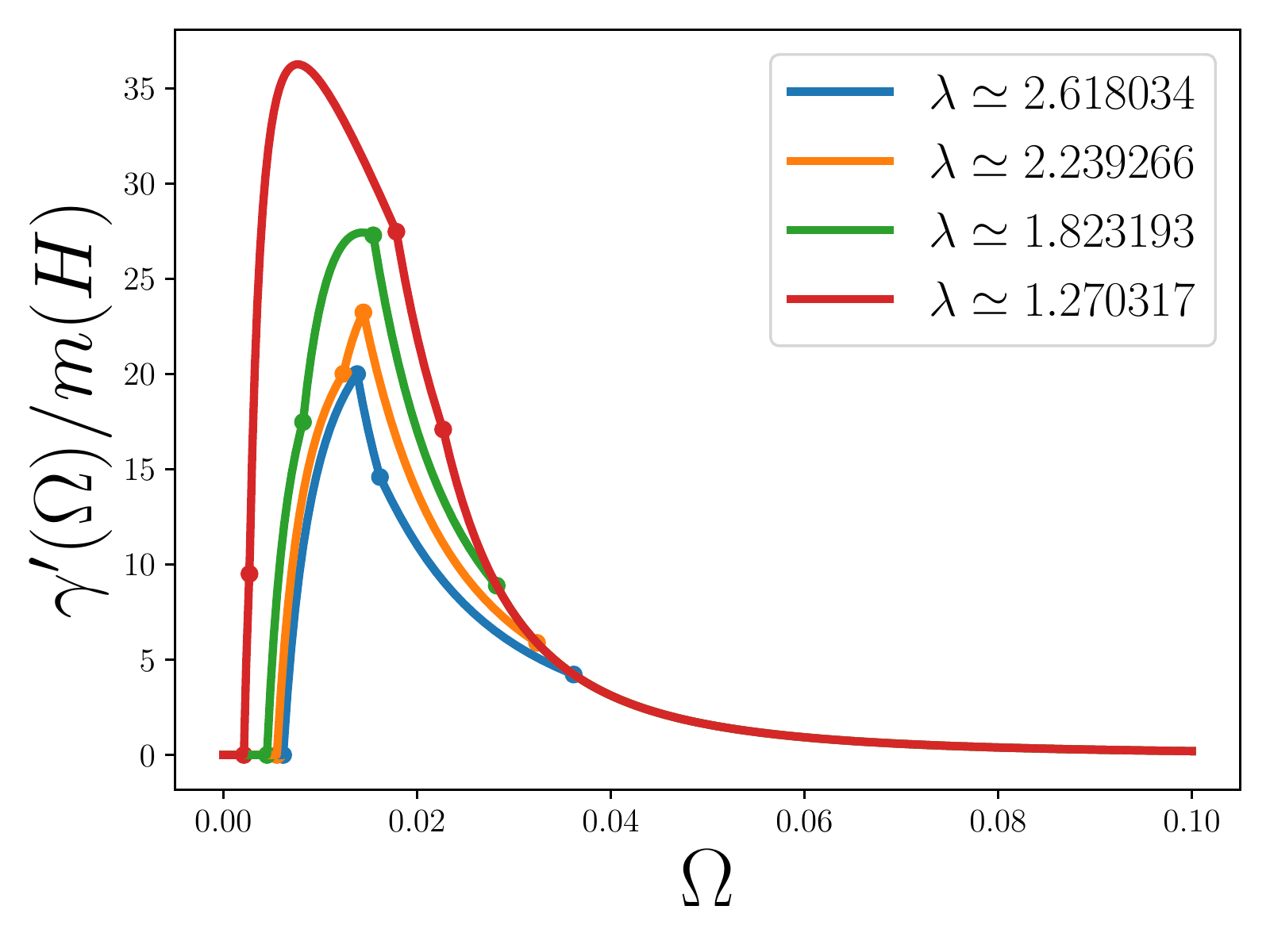}
\includegraphics[width = 6.6cm,bb= 0 0 461 346]{fig_pdf/4_3b.pdf}
\end{center}
\end{minipage}
\caption{
(a) Derivative $\gamma'(\Omega)$ of the normalized escape rate obtained using the formulas (\ref{eq:gamma1}) and (\ref{eq:gamma2}) plotted as a function of the noise strength. 
$\lambda$ denotes the instability defined in the local model. 
The dots in each curve represent the edges of each interval in the formulas (\ref{eq:gamma1}) and (\ref{eq:gamma2}). 
The area of the leak is set to $m(H)=10^{-4}$. (b) Numerically computed derivative from simulations of the cat map (same graph as Fig.~\ref{fig:fig5}(b)).
}
\label{fig:fig10}
\end{figure}
%%%%%%%%%%%%%%%%%%%%%%%%%%%%%%%%%%%%%%%%%
%%%%%%%%%%%%%%%%%%%%%%%%%%%%%%%%%%%%%%%%%
%%%%%%%%%%%%%%%%%%%%%%%%%%%%%%%%%%%%%%%%%

\subsection{Comparing local model and numerical results}
\label{sec:comparison}

Numerical analysis is now performed on the noisy cat map~(\ref{eq:perturbed_cat_map_noise}), in order to test the validity of the predictions obtained by the local model (see Fig.~\ref{fig:fig8}). 
We also examine how the local model explains the behavior of the derivative $\gamma'(H,\Omega)$ of the normalized escape rate obtained using the formulae (\ref{eq:gamma1}) and (\ref{eq:gamma2}) plotted as a function of the noise strength (see Fig.~\ref{fig:fig10}). 
%$\lambda$ denotes the instability defined in the local model. 
%The dots in each curve represent the edges of each interval in Eqs.~(\ref{eq:gamma1}) and~(\ref{eq:gamma2}). 
The area of the leak is set to $m(H)=10^{-4}$, as in 
the numerical results found in Fig.~\ref{fig:fig5}. 

Figure~\ref{fig:fig8}(a) and (b) respectively illustrate the case with $1 < \lambda \le 1 + \sqrt{2}$ and with $1+ \sqrt{2} < \lambda$.
In both cases, initial plateaux in the vicinity of $\Omega = 0$ are well reproduced by the local model. As indicated above, the existence of the plateau comes from the range  
$0<\Omega<l\left(1-1/\lambda\right)$ of the noise amplitude. The escape rate with noise coincides with that obtained in the noiseless case and thus is not affected by noise. 
It was observed in numerical results that the length of the plateau is reduced as the instability of the fixed point decreases. Such a tendency is not only reproduced in the model but can also be read from expressions~(\ref{eq:gamma1}) and~(\ref{eq:gamma2}). 
We have drawn vertical lines on the $\gamma(\Omega)/m(H)$ vs. $\Omega$ plots, to
 indicate the change of functional form in the local model: the plateau is followed by regions of linear ( and almost linear) growth in $\Omega$; for larger noise amplitudes, the 
 escape rate increases more slowly (to leading order$\sim c(l,\lambda)\Omega - \Omega^2$), until it saturates to the value of the
 area of the leak $m(H)$ for strong noise.
       
It is found that the estimate $m(\Theta_{1,\Omega}(H))$ for the escape rate in the local model exceeds the numerically obtained values of the escape rate $\gamma(H,\Omega)$ of the cat map over the full range of 
noise amplitudes examined. 
That may be due to the assumption in the local model  that the orbits are uniformly distributed in the whole available phase space, which  
is not necessarily valid. In particular, it tends to break down along the unstable manifold emanating from the periodic point at the center of the hole. 
Possibly for that reason, the curves predicted by the local model overestimate numerical results, and the agreement is only qualitative, at this point. 
Still, $m(\Theta_{1,\Omega}(H))$ overall proves an adequate qualitative descriptor for $\gamma(H,\Omega)$.      
%This would be because the assumption (A) is not valid to capture what happens in the noisy dynamics. 

Figure~\ref{fig:fig9} displays the orbit density around the hole after $n=100$ iterates of the noisy cat map, from a uniform initial density. 
The assumption (i) of subsection~\ref{sec:local_model}, that is that the orbits are uniformly distributed in the region $M \setminus H$ for $n \ge 0$,
 is apparently satisfied for large enough noise amplitude $\Omega$ (see Fig.~\ref{fig:fig9}(c)). Instead, it breaks down for smaller $\Omega$, as seen in Fig.~\ref{fig:fig9}(b).
In this case, the deterministic dynamics outplays noise in the evolution of densities. 
This signature of the dynamics around the fixed point provides some intuition for the periodic-orbit corrections to the area of the leak in the analytic expressions of the escape rate. 
The light-shaded region, most clearly seen in the noiseless simulation (see Fig.~\ref{fig:fig9}(a)), appears along the unstable manifold of the fixed point of the hole in Fig.~\ref{fig:fig9}(b). 
The observed inhomogeneity of the overall density of trajectories may account for the systematic overestimation of the escape rate by the local model, that relies on the 
assumption of uniformity, instead.

The response to noise depends on the instability of the fixed point contained in the hole, as observed in Fig.~\ref{fig:fig5}. 
This aspect can also be accounted for by the local model: as shown in Fig.~\ref{fig:fig10}, the weaker the instability, the larger the response to noise. Although the escape rate estimated from $m(\Theta_{1,\Omega}(H))$  
does not accurately predict the observed numerical results, profiles are qualitatively well reproduced. 
The initial sharp increase in the response may also be readily seen analytically by differentiating Eqs.~(\ref{eq:gamma1}) and~(\ref{eq:gamma2})  
%The formula describes the initial rapid increase and the peak in the curve $\gamma'(\Omega)/m(H)$ 
in the second and third noise ranges identified in the local model. 

We finally comment the behavior of $\gamma(H,\Omega)$ in the limit of $l\rightarrow 0$ and $\Omega \rightarrow 0$. 
If one lets $\Omega\rightarrow 0$ with $l$ fixed, the first case (weak noise regime) in Eqs.~(\ref{eq:gamma1}) and~(\ref{eq:gamma2}) applies, leading to the initial plateau in the 
$\gamma(H,\Omega)$ vs.\,$\Omega$ graph.
In the opposite case, that is, if one lets $l\rightarrow 0$ with $\Omega$ fixed, the last case (dominant noise) in Eqs.~(\ref{eq:gamma1}) and~(\ref{eq:gamma2}) appears. 
We therefore stress that formulae~(\ref{eq:gamma1}) and~(\ref{eq:gamma2}) still make sense in the limit $l\rightarrow 0$ and $\Omega\rightarrow 0$ when the ratio $l/\Omega$ is kept fixed.

%%%%%%%%%%%%%%%%%%%%%%%%%%%%%%%%%%%%%%%%%
%%%%%%%%%%%%%%%%%%%%%%%%%%%%%%%%%%%%%%%%%
\section{Summary and outlook}
\label{sec:summary}

We have investigated how the escape rate of the two-dimensional perturbed cat map behaves when uniformly-distributed noise is applied. 
%In order to normalize the issue, 
%we have limited 
The study has been focused on small enough areas of the leak, so as for the escape rate to feature a correction originating from the periodic orbit 
with the lowest period \cite{paar1997bursts,afraimovich2010hole,bunimovich2011place} inside the opening. 
As verified in subsection \ref{sec:numerical_verification}, an analytic estimate for the escape rate of the noiseless map is a good prediction for the numerical results. 
In particular, it was confirmed that the agreement improves as the area of the leak decreases, and the occasional deviation of the numerics from the theory could be attributed to the fact that other, relatively short periodic orbits, are contained in the hole surrounding the periodic point that belongs to the shortest cycle. 

In this setting, we have added noise to the system and observed its response. 
As mentioned in the introduction,  the cat map is uniformly hyperbolic, which leads to structural stability against smooth perturbations \cite{katok1997introduction}. 
Thus, in principle, one may expect a similar feature when the system is perturbed by noise, and yet, noise is a type of perturbation beyond the argument of structural stability in standard dynamical systems, which calls for investigation. 

Numerical simulations have revealed that a plateau indeed appears around the vanishing noise strength, meaning that the escape rate keeps constant in the presence of weak noise. We conclude that a kind of structural stability is at work, here. 

Remarkably, the length of the plateau increases with the instability $\Lambda_p$ of the periodic orbit contributing the largest correction to the escape rate. In addition, the response to noise, measured by the derivative of the escape rate with respect to the noise strength, becomes sharper with the decrease of $\Lambda_p$.

We have proposed a local model in the vicinity of the opening, in order to explain the numerical outcomes of the response to noise. We have analytically estimated the escape rate, and found that
the local model qualitatively captures the numerical results, and, in particular, it explains the existence of a plateau around the zero noise limit: the escape rate does not depend on the noise strength, when the latter is sufficiently small. 
%Remark that the right edge of the plateau is controlled by the eigenvalue in the stable direction, not by the one in the unstable direction. 
In addition, the analytic results account for the response to noise becoming sharper with the decrease of instability.

There appear five intervals in the formula, the edges of each interval being determined by 
the local eigenvalues associated with the shortest periodic orbit. 
It should be noted that the eigenvalues of stable and unstable directions appear mixed, meaning that the response profile is determined by the interplay between stable and unstable directions. 

The local model could be easily modified in order to reflect the tilted stable and unstable 
manifolds in the cat map and perturbed cat map as well. 
On the other hand, future work will involve further improvements to incorporate the non-uniformity of density evolution observed in Fig.~\ref{fig:fig9}, that we identify as 
%This non-uniformity would be 
a signature of deterministic dynamics. %not smeared out by noise. 
A theory encompassing different-shaped leaks and non-trivial noise distributions should also be pursued, with the intent of extending the present results to a more general
framework.

\section{Acknowledgements}
The authors are grateful to Kensuke Yoshida for 
enlightening discussions.
This work has been supported by JSPS KAKENHI Grants No.17K05583. 
DL is partially supported by the National Science Foundation of China, Grant No. 11750110416-1601190090.

\appendix
%%%%%%%%%%%%%%%%%%%%%%%%%%%%%%%%%%%%%%%%%
%%%%%%%%%%%%%%%%%%%%%%%%%%%%%%%%%%%%%%%%%
%%%%%%%%%%%%%%%     Appendix A   %%%%%%%%%%%%%%%%
%%%%%%%%%%%%%%%%%%%%%%%%%%%%%%%%%%%%%%%%%
%%%%%%%%%%%%%%%%%%%%%%%%%%%%%%%%%%%%%%%%%
\section{Alternative noisy evolution}
\label{app:alter_evol}

In order to crosscheck the estimates for the upper bound of the escape rate in the local model, let us follow an alternative approach 
for the convolution of dynamics and distributions, by
merging deterministic dynamics and noise into a single operation.
Apply the transfer operator to $P_{H,\Omega}(y)$, instead of to $\capchi_{M\left/H\right.}(x)$, as we
did in Eq.~(\ref{ThetaOm1}):
%\begin{linenomath}
\begin{eqnarray}
m\left(\Theta_{1,\Omega}(H)\right) &=& \int_M dy \, P_{H,\Omega}(y) \int_M dx\, \delta(y-f(x))\capchi_{M\left/H\right.}(x)  
\\ &=&  
 \int_M dx \, P_{H,\Omega}\left(f(x)\right) \capchi_{M\left/H\right.}(x)
\,.
\label{mTheta1}
\end{eqnarray}
%\end{linenomath} 
The iterated escape probability $P_{H,\Omega}\left(f(x)\right)$ is read as the probability for $f(x)$ to
be in the hole $H$, and it is  
obtained as in Eqs.~(\ref{eq:1-dimensional1}) and~(\ref{eq:1-dimensional2}):  
%\begin{linenomath}
\begin{equation}
P_\Omega\left(f(x)\right) = \frac{m\left(f^{-1}(H)\bigcap W_\Omega(x)\right)}{m\left(W_\Omega(x)\right)}
\,,
\end{equation}
%\end{linenomath}    
where $W_\Omega(x)=W_\Omega(X)\times W_\Omega(Y)$, with
%\begin{linenomath}
\begin{equation}
W_\Omega(X) = \left[X-\frac{\Omega}{2}, X + \frac{\Omega}{2}\right)
\,,
\end{equation}
%\end{linenomath}    
and $W_\Omega(Y)= W_\Omega(X)$. 
In the local model considered in section~\ref{sec:local_model}, the result is, for the $Y$-coordinate,
in the case $\Omega<\lambda l$,
%\begin{linenomath}
\begin{equation}
\hspace{-10mm}
P_{H,\Omega}^{(Y)}\left(f(x)\right) =
\left\{ 
\begin{array}{ll} 
\vspace{2mm}
\displaystyle 
0 & \hspace{5mm} Y<- (\lambda l + \Omega)/2, \\
%\frac{\lambda l + \Omega}{2} \\
\vspace{2mm}
\displaystyle 
\frac{Y}{\Omega} + \frac{\lambda l + \Omega}{2\Omega} & 
\hspace{5mm} 
-(\lambda l + \Omega)/2 \leq Y \leq -(\lambda l - \Omega)/2, \\
%-\frac{\lambda l + \Omega}{2}\leq Y \leq -\frac{\lambda l - \Omega}{2} \\
\vspace{2mm}
\displaystyle 
1 &  \hspace{5mm} 
-(\lambda l - \Omega)/2 < Y \leq (\lambda l - \Omega)/2, \\
%-\frac{\lambda l - \Omega}{2}< Y \leq \frac{\lambda l - \Omega}{2} \\
\vspace{2mm}
\displaystyle 
-\frac{Y}{\Omega} + \frac{\lambda l + \Omega}{2\Omega}  
&  \hspace{5mm} 
(\lambda l - \Omega)/2 < Y \leq (\lambda l + \Omega)/2,  \\
\vspace{2mm}
\displaystyle 
%\frac{\lambda l - \Omega}{2}< Y \leq \frac{\lambda l + \Omega}{2} \\
0 & \hspace{5mm} 
Y> (\lambda l + \Omega)/2. 
%Y> \frac{\lambda l + \Omega}{2}
\end{array}
\right.
\,
\end{equation}
%\end{linenomath}  
and similarly for the $X$-coordinate, where the $f^{-1}$ squeezes the length $l$ of the hole by a factor of $\lambda$, instead. 
At this point, Eq.~(\ref{mTheta1}), for the measure of the set of trajectories that escape in $n=1$ iterations, may be evaluated:
%\begin{linenomath}
%\begin{adjustwidth}{-\extralength}{0cm}
%\begin{flushleft}
\begin{eqnarray}
\nonumber
\hspace{-15mm}
m\left(\Theta_{1,\Omega}(H)\right) &=& \int_M dx \, P_{H,\Omega}\left(f(x)\right) \capchi_{M\left/H\right.}(x) \\ \nonumber
&=& \int_{-1/2}^{-l/2}dX\int_{-1/2}^{1/2} dY  P_{H,\Omega}\left(f(x)\right) +    \int_{-l/2}^{l/2}dX\int_{-1/2}^{-l/2} dY  P_{H,\Omega}\left(f(x)\right) \\ \nonumber
&+&  \int_{-l/2}^{l/2}dX\int_{l/2}^{1/2} dY  P_{H,\Omega}\left(f(x)\right)  +  \int_{l/2}^{1/2}dX\int_{-1/2}^{1/2} dY  P_{H,\Omega}\left(f(x)\right) \\ \nonumber
&=&  2\int_{l/2}^{1/2}dX\int_{-1/2}^{1/2} dY  P_{H,\Omega}\left(f(x)\right) %+  2\cancel{\int_{-l/2}^{l/2}dX \int_{l/2}^{1/2}dY P_{H,\Omega}\left(f(x)\right)}  
\\ \nonumber
%2\int_{-l/2}^{l/2}dX\int_{l/2}^{1/2} dY  P_{H,\Omega}\left(f(x)\right) \\ \nonumber
&=& 2\left[  \int_{l/2}^{(\lambda l-\Omega)/2}dY +   \int_{(\lambda l-\Omega)/2}^{(\lambda l+\Omega)/2}dY \left(\frac{\lambda l}{2\Omega}-\frac{Y}{\Omega}\right)\right]\frac{l}{\lambda} \\ 
&=& l^2\left(1-\frac{1}{\lambda}\right)
%2\cancel{\int_{-l/2}^{l/2}dX \int_{l/2}^{1/2}dY P_{H,\Omega}\left(f(x)\right)} 
%= l^2\left(1-\frac{1}{\lambda}\right)
\,,
\label{mTheta1ev}
\end{eqnarray}
%\end{flushleft}
%\end{adjustwidth}
%\end{linenomath} 
in the regime of smallest $\Omega$.

%%%%%%%%%%%%%%%%%%%%%%%%%%%%%%%%%%%%%%%%%%
%%%%%%%%%%%%%%%%%       Fig. A1         %%%%%%%%%%%%%%%
%%%%%%%%%%%%%%%%%%%%%%%%%%%%%%%%%%%%%%%%%%
%\begin{figure}
%\begin{minipage}{1.0\hsize}
%\begin{center}
%\includegraphics[width = 5.0cm,bb= 0 0 461 346]{fig_pdf/POmegaX.pdf}
%\includegraphics[width = 5.0cm,bb= 0 0 461 346]{fig_pdf/POmegaFX.pdf}
%\includegraphics[width = 5.0cm,bb= 0 0 461 346]{fig_pdf/POmegaFY.pdf}
%%\includegraphics[clip,width=27mm]{./fig_pdf/4_12a.pdf}
%%\includegraphics[clip,width=27mm]{./fig_pdf/4_12b.pdf}
%%\includegraphics[clip,width=27mm]{./fig_pdf/4_12c.pdf}
%\end{center}
%\end{minipage}
%\caption{
%\caption{(a) $P_{H,\Omega}(X)$, (b) $P_{H,\Omega}\left(f(X)\right)$, (c) $P_{H,\Omega}\left(f(Y)\right)$, with $\lambda=2$, $\Omega=0.2$, $l=1$. }
%}
%\label{fig:fig9}
%\end{figure}
%
\begin{figure}[h]
\begin{minipage}{1.0\hsize}
\begin{center}
\includegraphics[width = 6.0cm,bb= 0 0 461 346]{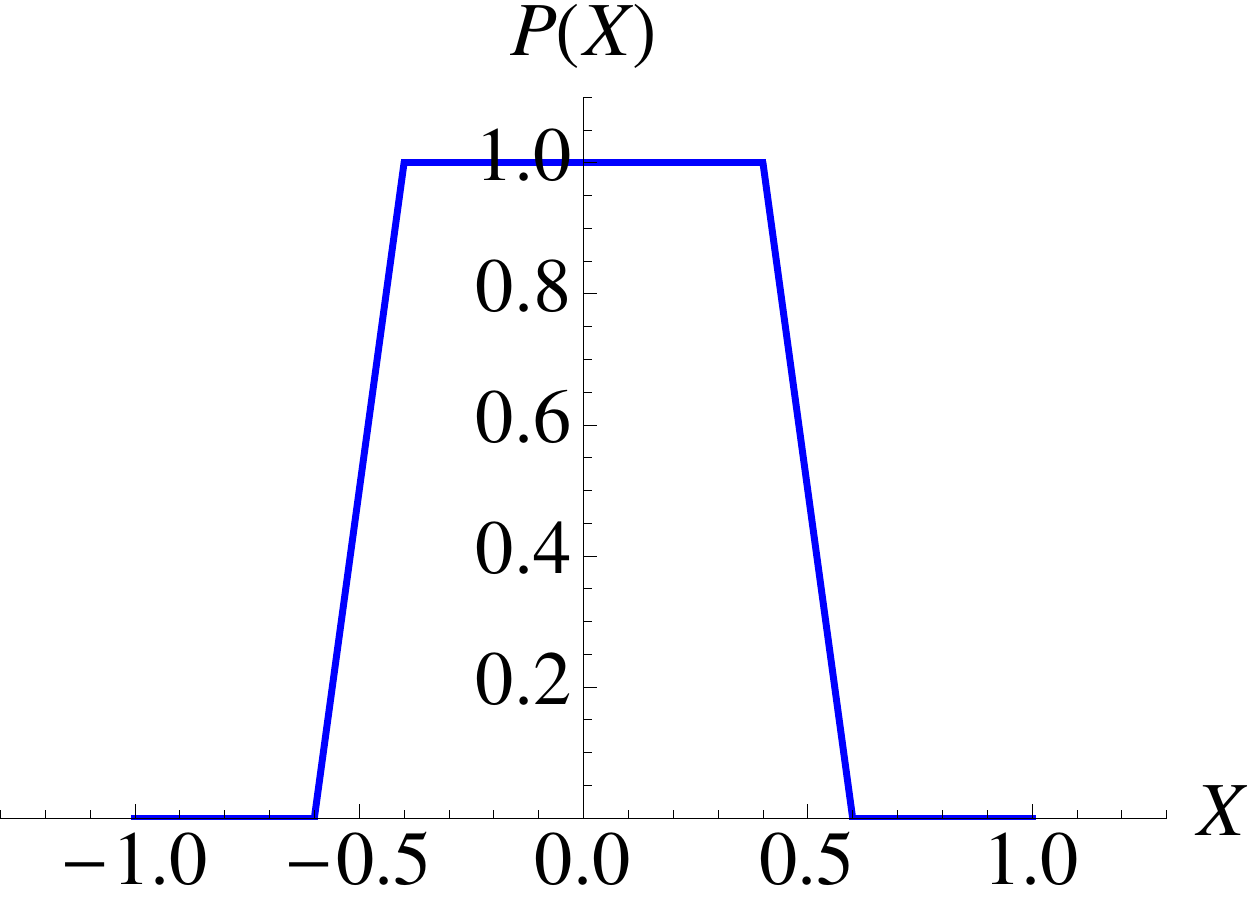}
\includegraphics[width = 6.0cm,bb= 0 0 461 346]{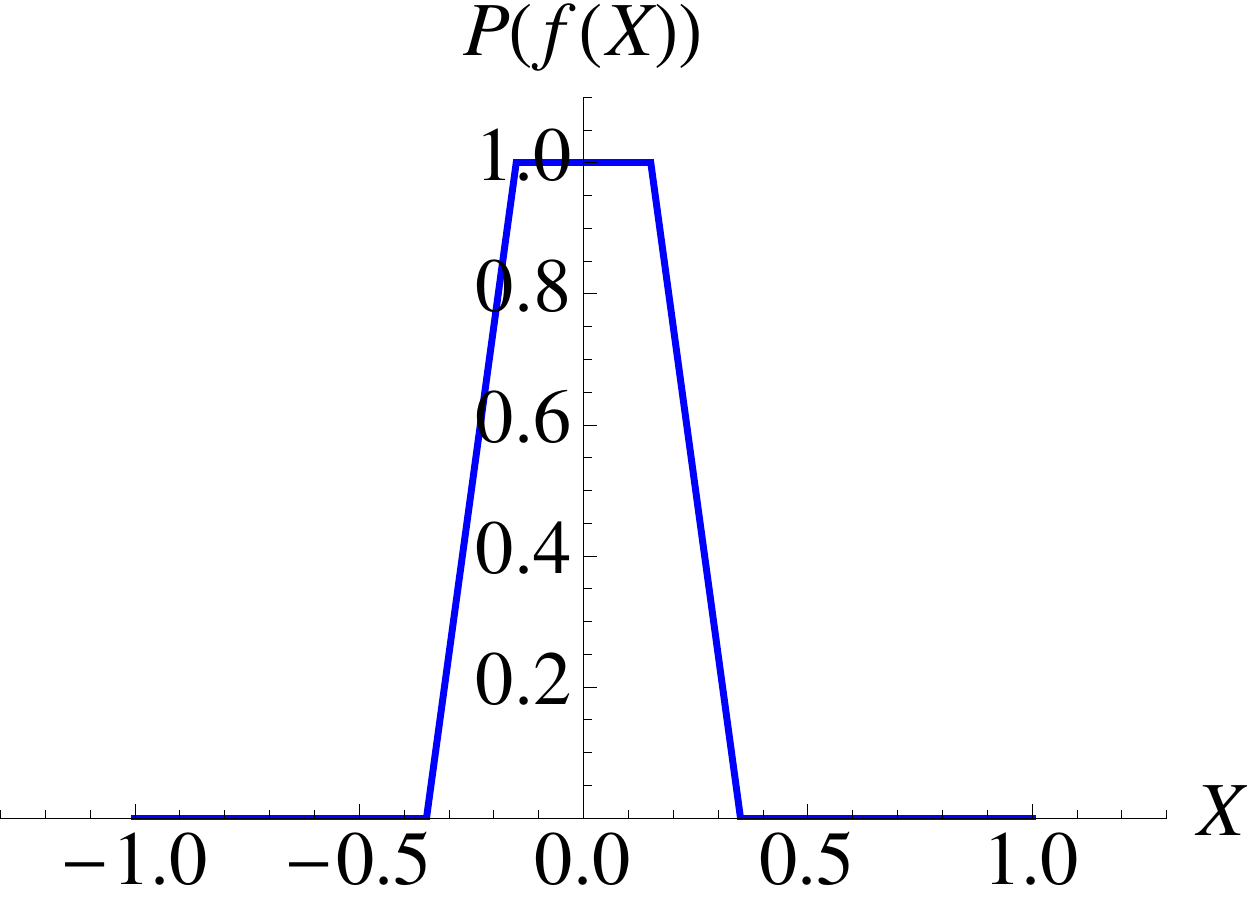}
\includegraphics[width = 6.0cm,bb= 0 0 461 346]{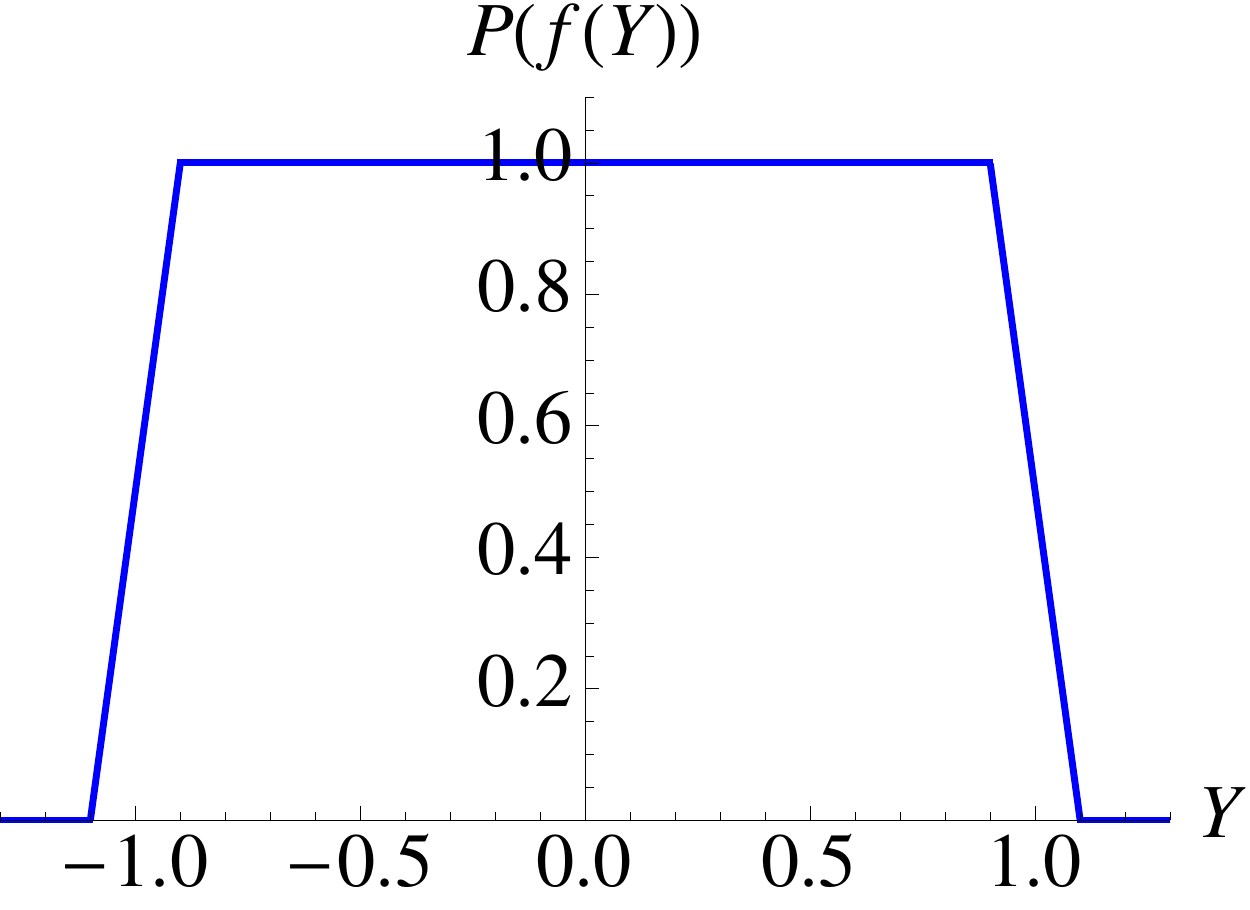}
\end{center}
\end{minipage}
\caption{Probability densities: (a) $P_{\rm H,\Omega}^{\rm (X)}(x)$, (b) $P_{\rm H,\Omega}^{\rm (X)}\left(f(x)\right)$, (c) $P_{\rm H,\Omega}^{\rm (Y)}\left(f(x)\right)$, with $0<\Omega<l(1-1/\lambda)$. }
%\vspace{30mm}
\label{fig:pxyb_lam}
\end{figure}
\bibliographystyle{model3-num-names}
\bibliography{reference}

\ifboyscout
\newpage
\input{../../blog/flotperturb}
\fi

\end{document}